\documentclass[a4paper,11pt]{quantumarticle}
\pdfoutput=1

\usepackage[utf8]{inputenc}
\usepackage[english]{babel}
\usepackage[T1]{fontenc}
\usepackage{amsmath}
\usepackage{hyperref}
\usepackage{booktabs}
\usepackage{multirow}
\usepackage{subfigure}
\usepackage{caption}
\usepackage{amsmath}
\usepackage{amsthm}
\usepackage{physics}
\usepackage{amssymb}
\usepackage{cite}
\usepackage{soul}
\usepackage[export]{adjustbox}
\usepackage{graphicx}
\usepackage{tabularx}
\usepackage[commentsnumbered,ruled,vlined]{algorithm2e}
\usepackage[normalem]{ulem}

\DeclareUnicodeCharacter{2212}{-}

\newtheorem{definition}{Definition}
\DeclareMathOperator*{\argmax}{arg\,max}

\usepackage{pifont}

\begin{document}

\title{Energy risk analysis with Dynamic Amplitude Estimation and Piecewise Approximate Quantum Compiling}

\author{Kumar Ghosh}
\email{kumar.ghosh@eon.com}
\orcid{0000-0002-4628-6951}
\affiliation{E.ON Digital Technology GmbH, Essen, Germany}

\author{Kavitha Yogaraj}
\affiliation{IBM Quantum, IBM Research, India}

\author{Gabriele Agliardi}
\orcid{0000-0002-1692-9047}
\affiliation{IBM Italia, Milan, Italy}

\author{Piergiacomo Sabino}
\affiliation{E.ON SE, Essen, Germany}
\affiliation{University of Helsinki, Finland, Department of Mathematics and Statistics}
\orcid{0000-0003-2072-2353}

\author{Marina Fernández-Campoamor}
\affiliation{E.ON Digital Technology GmbH, Essen, Germany}

\author{Juan Bernabé-Moreno}
\affiliation{IBM Research Europe, Dublin, Ireland}

\author{Giorgio Cortiana}
\affiliation{E.ON Digital Technology GmbH, Essen, Germany}

\author{Omar Shehab}
\affiliation{IBM Quantum, IBM Thomas J Watson Research Center, Yorktown Heights, NY, USA}

\author{Corey O'Meara}
\email{corey.o'meara@eon.com}
\affiliation{E.ON Digital Technology GmbH, Essen, Germany}

\maketitle

\begin{abstract}
We generalize the Approximate Quantum Compiling algorithm into a new method for CNOT-depth reduction, which is apt to process wide target quantum circuits. Combining this method with state-of-the-art techniques for error mitigation and circuit compiling, we present 
a 10-qubit experimental demonstration of Iterative Amplitude Estimation on a quantum computer. The target application is the derivation of the Expected Value of contract portfolios in the energy industry.

In parallel, we also introduce a new variant of the Quantum Amplitude Estimation algorithm which we call Dynamic Amplitude Estimation, as it is based on the dynamic circuit capability of quantum devices. The algorithm achieves a reduction in the circuit width in the order of the binary precision compared to the typical implementation of Quantum Amplitude Estimation, while simultaneously decreasing the number of quantum-classical iterations (again in the order of the binary precision) compared to the Iterative Amplitude Estimation. The calculation of the Expected Value, VaR and CVaR of contract portfolios on quantum hardware provides a proof of principle of the new algorithm.

\end{abstract}

\section{Introduction}\label{Sec:Introduction}

\begin{figure*}
\centering \footnotesize
\input{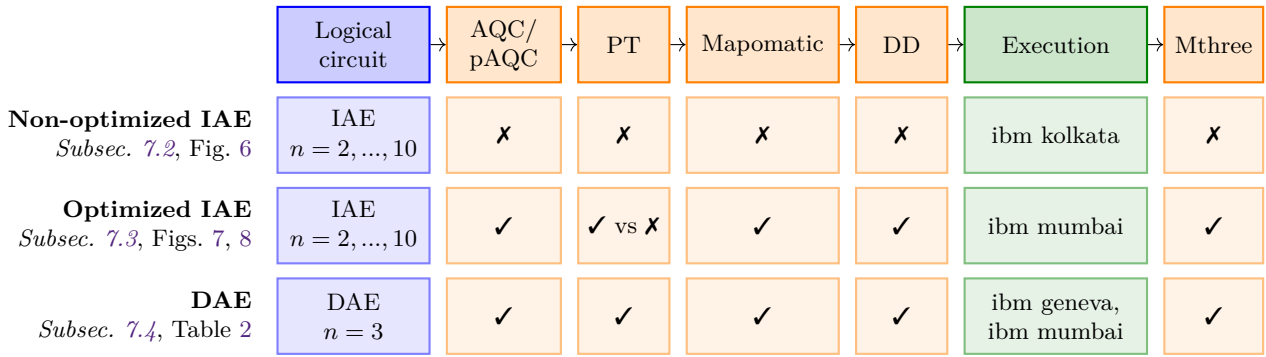}
\caption{The first row describes the workflow for the production and execution of the quantum circuits used for our estimation. Specifically, the first blue box represents the generation of the logical circuits, whose general method is described in Sec.~\ref{A quantum approach}. The green box represents the circuit execution on quantum hardware. The orange boxes represent error mitigation and circuit optimization techniques that reduce the effect of hardware noise: Pauli Twirling (PT), mapomatic, Dynamical Decoupling (DD) and mthree are state-of-the-art techniques described in Subsec.~\ref{subsec:stateoftheart}; Approximate Quantum Compiling (AQC) is presented in Subsec.~\ref{subsec:aqc}, and extended into the new variant of piecewice AQC (pAQC) in Subsec.~\ref{subsec:PAQC}.
The other lines of the diagram represent the core experiments of the paper. In the `logical circuit' column, we indicate whether Iterative Amplitude Estimation (IAE) or the new Dynamic Amplitude Estimation (DAE) is employed, and the number $n$ of qubits in the input distribution. In the orange columns, we indicate which optimization techniques were used: pAQC is employed in lieu of AQC for $n \geq 8$. Finally, in the execution column we provide the name of the IBM Quantum backend.}\label{fig:summary}
\end{figure*}

The computation of risk measures can be made more time-efficient by resorting to quantum computers, and specifically to the Quantum Amplitude Estimation (QAE) algorithm, but current noisy quantum hardware limits short-term possibilities of real-world applications.
We demonstrate three novel contributions in this work.
First, we propose a variant of the Approximate Quantum Compiling algorithm (AQC), called Piecewise Approximate Quantum Compiling (pAQC). Like the original AQC, our method is designed to approximate circuits in order to reduce their depth, but the novel variant is more suited for deep circuits.
Second, we provide a 10-qubit experimental demonstration of amplitude estimation in quantum hardware, using the Iterative Amplitude Estimation (IAE) in combination with state-of-the art error mitigation and circuit optimization techniques as well as the newly introduced pAQC. We showcase the execution of a 10-qubit amplitude estimation circuit on a quantum computer.
Finally, we introduce the Dynamic Amplitude Estimation (DAE), an upgraded version of QAE, leveraging on the dynamic circuit capability recently introduced on IBM Quantum devices~\cite{pressrelease_dynamic_2021}, thus reducing the circuit width and the number of quantum-classical iterations, in comparison with the previously known equivalent techniques.

Our target application is the calculation of statistical quantities like the Expectation, VaR, and CVaR of the delta gross margin, namely the random variable representing the difference between the forecasted value of a contract portfolio, and its stochastic economic outcome. %
A faster calculation of such metrics would lead to a real-time planning and decision making, finer risk diversification, as well as more frequent risk assessments for the negotiation of hedging contracts, with radical impacts on risk management in the industry sector. 
To compute the above risk statistics we use one of the fundamental quantum algorithms called QAE, which achieves a quadratic speedup over classical Monte Carlo (MC) simulation, e.g. in financial services for the purpose of option pricing\cite{stamatopoulos2020option, rebentrost_quantum_2018} and risk analysis~\cite{woerner2019quantum, egger_credit_2021, laudage_calculating_2022, dri_more_2023}. The canonical version of QAE is a combination of Quantum Phase Estimation (QPE) \cite{Kitaev1996QuantumMA} and Grover’s Algorithm \cite{Grover}. The size of the QAE-circuit grows rapidly with the size of encoded probability distribution and the expected accuracy of the  result, therefore it needs a large amount of quantum resources for the implementation.
Aiming to reduce the required resources, several QAE variants have been introduced in literature, in which QPE is replaced by ad-hoc classical workloads. Early attempts showed either lack of rigorous proof \cite{suzuki_amplitude_2020, wie_simpler_2019}, or high constants rendering the algorithm unsuitable for practical applications \cite{aaronson_quantum_2019}.
The latest developments include Iterative Amplitude Estimation (IAE)~\cite{Grinko2021} and ChebQAE~\cite{rall_amplitude_2022}, which have both a rigorous proof for the quadratic speed-up (up to a multiplicative $\log(\log(\epsilon^{-1}))$ factor) and much improved constants.

\paragraph{}
The outline of the article is as follows. Sec~\ref{sec:summary-res} contains a summary of our results. In Sec.~\ref{Sec:Background}, we begin with an overview of the energy contract risk analysis and the relevant risk measures. In the following Sec.~\ref{A quantum approach}, we summarize the workflow, main steps, and the mathematical preliminaries for QAE-inspired risk analysis. In the next Sec.~\ref{sec:DAE}, we introduce the novel Dynamic Amplitude Estimation. In Sec.~\ref{Sec: Experimental demonstration}, we compute the statistics of the random variable, first using the qasm simulator and then using the quantum computer, with QAE and DAE. The circuit is altered to reduce depth and optimize for the quantum hardware, via multiple techniques: optimal qubit mapping, dynamical decoupling with mapomatic, error mitigation with mthree, Pauli Twirling, and approximate quantum compiler (AQC). Specifically we propose a variant of AQC that operates on wide circuits, the pAQC. In Sec.~\ref{Sec:Discussion}, we highlight some important points of our development and the advantages over its non-dynamical counterparts.  Finally, in Sec.~\ref{Sec: Conclusion} we offer some concluding remarks.

Fig.~\ref{fig:summary} shows how the different error mitigation and circuit optimization techniques combine with IAE and DAE in the experiments.

\section{Result summary}\label{sec:summary-res}
This section summarizes our three main contributions.

\paragraph{Piecewise Approximate Quantum Compiling.}
Current hardware capabilities constrain the maximal depth of executable circuits, before the noise prevails. Therefore, approximating a circuit with an alternate one having lower depth, proves to be extremely beneficial. Approximate Quantum Compiling (AQC)~\cite{AQC} provides very good results, but unfortunately suffers from a barren plateau effect on wide circuits, and cannot be employed beyond a threshold of $7$ or $8$ qubits, in our experience. Consequently, we contribute with a variant that we call piecewise AQC (pAQC), designed for wide circuits.

\paragraph{A 10-qubit demonstration with IAE, noise mitigation and circuit approximation.} Literature lacks demonstrations of QAE and related techniques beyond very few qubits on quantum computer, with the exception of Ref.~\cite{dri_more_2023} that unfortunately does not give details on the effect of errors over estimations. By combining state-of-the-art capabilities in error mitigation and circuit approximation~\cite{mapomatic, dynamicdecoupling, m3errormitigation, geller2013efficient}, we are able to estimate the Expectation through circuits up to $7$ data qubits.%
Additionally, by resorting to pAQC, we estimate the Expectation of distributions up to $10$ qubits, with similar error rates. This achievement is far beyond any witnessed result, at the best of the authors' knowledge.

\paragraph{Dynamic Amplitude Estimation.}
The quantum-classical iterative methods, including IAE used above, require multiple subsequent hardware calls, and preclude the possibility to generate all circuits in a single preprocessing phase, thus adding communication overhead, depth unpredictability and warm-start complexities.
At the same time, a recent hardware capability, namely dynamic circuits, offers new levers for circuit depth optimization. Correspondingly, a variant of QPE (here called DPE) was developed to reduce the number of qubits, and therefore the gate error propagation, of the original QPE \cite{dynamic_circuits}. %
In this work, we introduce the Dynamic Amplitude Estimation (DAE) in which the qubit-intensive form of original QPE is replaced by a phase estimation circuit that exploits dynamic circuits. %
The DAE algorithm, compared to QAE, reduces the required number of qubits (from $m$ to 1, for $m$ significant bits), and thus the gate error propagation. Additionally, re-using the same qubit for all measurements implies increased consistency in readout and CNOT errors \cite{oloan_iterative_2010}. Compared to iterative methods such as IAE, DAE reduces the required number of quantum-classical iterations from $\mathcal O (m)$ to $1$, allowing for contemporary submission of all jobs. These facts lower the bar to apply QAE in practice. At the same time, DAE has a rigorous proof, which is a mandatory requirement in regulated environments.

The techniques introduced here are relevant beyond the context of energy risk analysis. Indeed, pAQC is beneficial for the reduction of the depth of virtually any circuit. Similarly, DAE is applicable to all fields where QAE methods are employed: primarily for Monte Carlo Integration~\cite{montanaro_quantum_2015} with implications in finance~\cite{stamatopoulos2020option, woerner2019quantum} and physics~\cite{knill_optimal_2007, agliardi_quantum_2022}, as well as for optimization~\cite{kaneko_linear_2021} and machine learning~\cite{wiebe_quantum_2016, kerenidis_q-means_2018}.

\section{Energy portfolio risk analysis} \label{Sec:Background}
In the gas industry, standard contracts for private or industrial customers normally entail fixed unitary prices, and do not include any volume constraints. On the other side, the gas demand of households or heating has a strong dependency on gas volumes and weather variables, typically the temperature, that in turn implies variations of prices in the gas trade market. In other words, the gas supplier takes the risk of volume (and price) deviations from the projected load profile of the customer.

As a consequence, risk managers compute the fair value and the financial exposure deriving from the entire weather-related portfolio, as well as of the individual contracts it consists of. To this end, they rely on a joint stochastic model  for the gas prices and temperatures, and perform extensive and time-consuming Monte Carlo simulations~\cite{Glass2004} to estimate such statistics.
Many classical models and approaches are available in the literature which encompass at the same time the joint temperature-gas evolution, the Monte Carlo simulation and the risk analysis (see for instance Benth and Benth~\cite{BenthBenth05, BenthBenth07}, Cucu et al~\cite{cucu_et_al16}, Sabino and Cufaro Petroni~\cite{cs20_2}).

A key economic value in an energy contract portfolio is the change in gross margin $\Delta GM$, defined as follows.
Consider a simplified weather-related portfolio which depends on gas and temperature. These two variables are called \emph{underlyings}. For simplicity, we can consider three gas European markets, Germany, UK, and Italy, and one temperature location relative to each country, Berlin, London, and Rome, respectively. 
The portfolio we focus on, only consists of supply contracts, based on which the customer can nominate gas volumes at an agreed sales price, denoted as \emph{asp}. These contracts are then implicitly temperature dependent: indeed,  the customers' demand is modeled by a volume function ($V(T)$) of the temperature.

In this setting, the delta gross margin is defined as the (unknown random) difference between the net random sales less the random costs at a certain time future time $t$ and the planned, therefore known, sales minus cost at the same future time.
Formally, let $i=1,2,3$ be an index over the weather stations, and $j=1,...,365$ an index over the time slices in the horizon. Accordingly, the portfolio consists of three contracts each dependent on its relative weather station. In addition, for simplicity, we assume that each station is located in a specific gas market, namely we consider three gas markets.
Denote by $T_{ij}$ the random temperature at time $j$ of the weather station $i$ generated by a suitable Monte Carlo model, and by $\tau_{ij}$ the corresponding \emph{season-normal}, that is the daily expectation of each temperature station. Similarly, denote by $G_{ij}$ the day-ahead gas price at time $j$ of the market $i$. Then, the change in gross margin of the $i$th contract in the portfolio is %
\begin{equation}\label{eq:deltaGM}
\Delta GM_i = \sum_{j}(V(T_{ij}) - V(\tau_{ij})) (asp_i - G_{ij})).
\end{equation}
Of course, the change in gross margin of the portfolio is the sum of that of its contracts.
In this study we take as an input a set of $\Delta GM$, generated for multiple contracts and times.
We investigate and refine the calculation of the Expected Value, the Value at Risk (VaR) and the conditional Value at Risk (CVaR) of $\Delta GM$ with the QAE and variants.

\section{Quantum approach for risk analysis: mathematical preliminaries}\label{A quantum approach}
\begin{figure*}[tp]
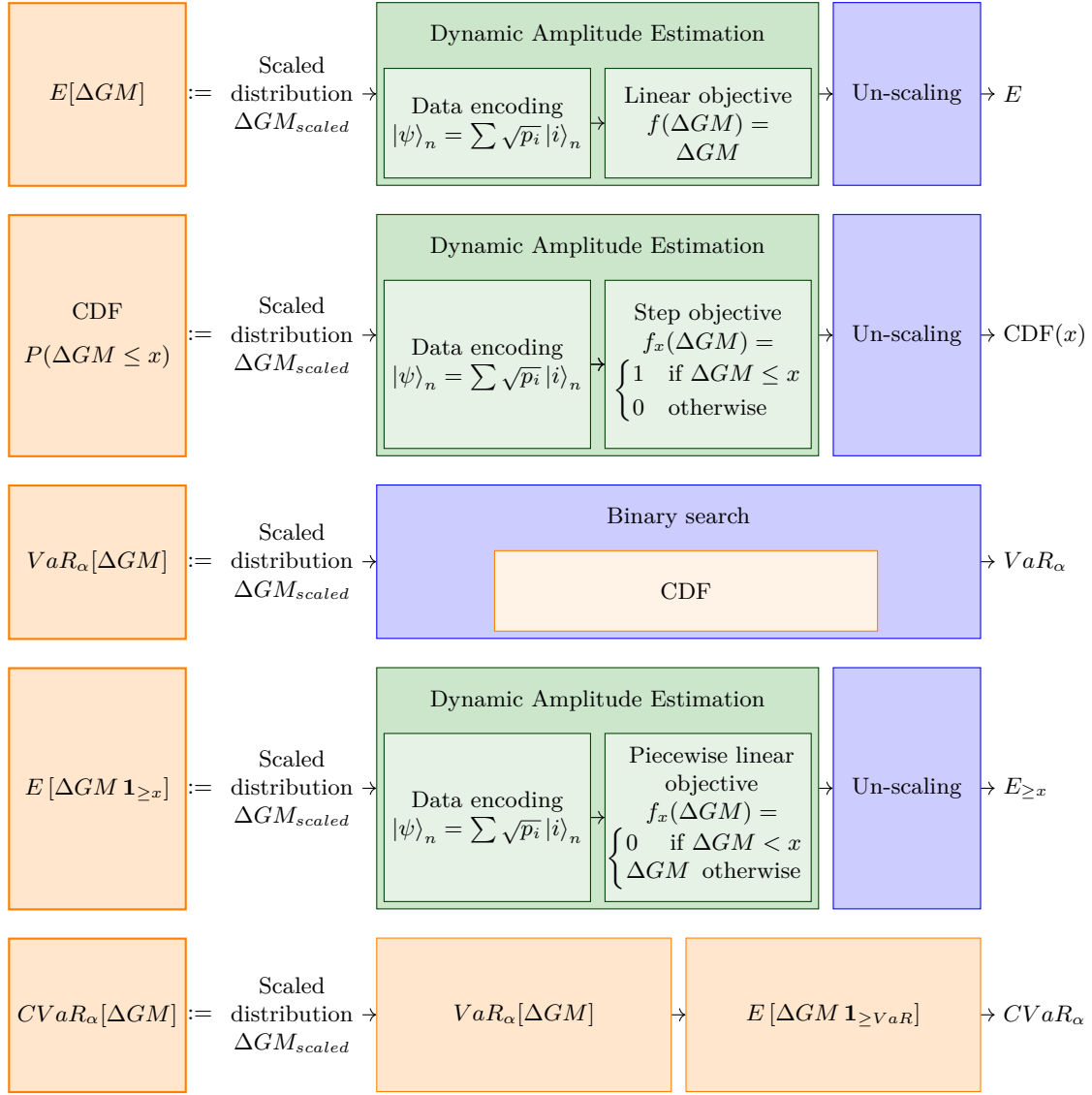

    \centering \footnotesize
    \include{tex_files/flow_chart}
    \caption{Flowcharts for the calculation of the relevant statistics, from top to bottom: Expectation ($E$), Cumulative Density Function (CDF), Value at Risk ($VaR$), Expectation on a subdomain $\{ \Delta GM \geq x\}$ ($E_{\geq x}$), Conditional Value at Risk ($CVaR$). Green boxes represent quantum workloads, while blue boxes are classical workloads. Some metrics are calculated relying upon other ones: specifically, the $VaR$ performs a (classical) binary search over the CDF, which is in turn calculated via a quantum-classical algorithm; similarly, the $CVaR$ is the estimation of the Expected Value on a subdomain determined by the $VaR$, where both the $VaR$ and the Expected Value are determined via quantum-classical algorithms.}
    \label{Flow_chart}
\end{figure*}

In this section we describe a quantum method for energy contract portfolio risk analysis. A summary is described by the flowchart in Fig.~\ref{Flow_chart}.

\subsection{The framework of Quantum Amplitude Estimation for risk analysis}\label{subsec: Quantum Amplitude Estimation} 
Let us begin with a short introduction of Quantum Amplitude Estimation (QAE), which is the main quantum algorithm we use in this work. Suppose a quantum operator $\mathcal{A}$ acts on an ($n+1$)-qubit state $\ket{ 0 }_{n+1}$ and produces the following state
\begin{equation}
\mathcal{A} \ket{ 0 }_{n+1} = \sqrt{1-a} \ket{ \psi_0 } _n \ket{ 0 } + \sqrt{a} \ket{ \psi_1 } _n \ket{ 1 }. \label{QAE_equation}
\end{equation}
Then the QAE~\cite{brassard2002quantum} approximates the amplitude $a$ in the above equation \eqref{QAE_equation}, with an estimate $\tilde{a}$.
The error satisfies $\abs{a - \tilde{a}} = \mathcal{O}(M^{-1})$, where $M$ is the number of applications of $\mathcal{A}$,
with probability of at least $8/\pi^2$ \cite{ woerner2019quantum}. The algorithm then extends to any desired confidence level. Therefore a quadratic speed up in the convergence rate is obtained, compared to classical Monte Carlo methods, for which the error is $\mathcal O (M^{-1/2})$, where $M$ is the number of Monte Carlo simulations this time.

The idea of QAE and variants, is based on the possibility to construct a Grover operator $\mathcal Q$ based on $\mathcal A$, whose eigenvectors are related to the desired value $a$ by a trigonometric relationship. Afterwards, the Quantum Phase Estimation is applied to efficiently retrieve the (cosine of) the eigenvectors, and therefore $a$.

Recently, QAE was applied to credit risk analysis and option pricing \cite{stamatopoulos2020option, woerner2019quantum}. The risk analysis with QAE aims at estimating statistics of a random variable $X$, by formulating the problem in such a way that the desired statistics are the amplitudes in Eq.~\eqref{QAE_equation}.

In order to do so, the first step requires to prepare a quantum state $\ket{ \psi } _n$ that represents $X$. More explicitey, let $X$ be valued in $\lbrace 0, 1, ..., N-1 \rbrace$ with $N=2^n$, $n$ being the number of qubits, and let $p_i$ be the probability of $X=i$. The encoding operator is $\mathcal{R}$ acting as:
\begin{equation}\label{eq:dataloading}
\mathcal{R} \ket{ 0 } _n = \ket{ \psi } _n = \sum_{i=0}^{N-1} \sqrt{p_i} \ket{ i }_n
\end{equation}
with $i \in \lbrace 0, 1, ..., N-1 \rbrace$.
Given any relevant quantity in risk analysis, represented by a classical function $f$:
\begin{equation}
f : \lbrace 0, 1, ..., N-1 \rbrace \rightarrow [0, 1],
\end{equation}
suppose we are able to construct a corresponding quantum operator $\mathcal F$ such that
\begin{equation}\label{eq:operatorF}
\mathcal F \ket{ i } _n \ket{ 0 } = \ket{ i } _n \left[ \sqrt{1 - f(i)} \ket{ 0 } + \sqrt{f(i)} \ket{ 1 }\right],
\end{equation}
for all $i \in \lbrace 0, 1, ..., N-1 \rbrace$. Then 
\begin{equation}
\begin{split}
\mathcal F \ket{ \psi } _n \ket{ 0 } =&
\sum_{i=0}^{N-1} \left(\sqrt{1 - f(i)}\sqrt{p_i}\right) \ket{ i } _n \ket{ 0 }\\
&+ \sum_{i=0}^{N-1} \left(\sqrt{f(i)} \sqrt{p_i}\right) \ket{ i } _n \ket{ 1 }.
\end{split}
\end{equation}

The probability of measuring $\ket{1 }$ in the rightmost qubit is equal to the \textit{Expectation} of $f$:
\begin{equation}
\sum_{i=0}^{N-1}  f(i) \, p_i = E \left[ f(X) \right].
\end{equation}
and the QAE machinery applies to obtain said estimation efficiently. We need to show that our metrics of interest, namely the Expected Value, the Value at Risk, and the Conditional Value at Risk, can be derived from the Expectation of some function $f$.

Let us now move to the \textit{Value at Risk}. For a confidence level $\alpha \in [0,1]$,
$VaR_\alpha (X)$ is defined as the smallest value of $X \in  \lbrace 0, 1, ..., N-1 \rbrace$
such that $\mathbb{P}[X \leq x] \geq (1- \alpha)$, namely
\begin{equation}
{VaR}_{\alpha}(X) = \inf \lbrace x \mid \mathbb{P}[X \leq x] \geq 1 - \alpha \rbrace.
\end{equation}
To find $VaR_\alpha (X)$ with QAE, for all $l \in  \lbrace 0, 1, ..., N-1 \rbrace$, we define a step function
\begin{equation} \label{eq:f-VaR}
f_l (i) =
\begin{cases}
1 & \text{if } i \leq l, \\
0 & \text{otherwise.} 
\end{cases}
\end{equation}
Corresponding to $f_l$ we further define an operator $\mathcal F_l$ such that
\begin{equation}\label{eq:Fl}
\mathcal F_l \ket{ i } _n \ket{ 0 } = \ket{ i } _n \ket{ f_l(i) },
\end{equation}
so that
\begin{equation}\label{eq:Fl-sum}
\mathcal F_l  \ket{ \psi } _n \ket{ 0 }  = \sum_{i=l+1}^{N-1}\sqrt{p_i} \ket{ i } _n \ket{ 0 } + \sum_{i=0}^{l}\sqrt{p_i} \ket{ i } _n \ket{ 1 }.
\end{equation}
After implementing QAE, the probability of getting $\ket{ 1 }$ is equal to 
\begin{equation}
\sum_{i=0}^{l} p_i = \mathbb{P} [X \leq l],
\end{equation}
that is the CDF of $X$. Therefore $VaR_\alpha (X)$, for a given confidence level $\alpha$, can be retrieved as the smallest $l=l_\alpha $ and can be calculated by means of bisection search.

Finally, let us address the \textit{conditional Value at Risk} $CVaR_\alpha (X)$, defined as
\begin{equation}
{CVaR}_{\alpha}(X) = E\left[ X \mid X \geq VaR_{\alpha}(X) \right] .
\end{equation}
In words, it is the conditional Expectation of $X$ restricted to $\lbrace 0, 1, ..., l_\alpha \rbrace $, where $l_\alpha = VaR_\alpha (X)$ is defined as before. Then, to estimate $CVaR_\alpha (X)$, we define
\begin{equation}\label{eq:f-CVaR}
f(i) = \frac{i}{l_\alpha} f_{l_\alpha} (i)
\end{equation}
and the associated operator $\mathcal F$ such that 
\begin{equation}
\begin{split}
&\mathcal F \ket{ \psi } _n \ket{ 0 } =\\
&\left( \sum_{i=l_\alpha +1}^{N-1}\sqrt{p_i} \ket{ i } _n
+ \sum_{i=0}^{l_\alpha}\sqrt{1- \frac{i}{l_\alpha}} \sqrt{p_i} \ket{ i } _n  \right) \ket{ 0 }\\
& \quad\quad + \sum_{i=0}^{l_\alpha} \sqrt{ \frac{i}{l_\alpha} } \sqrt{p_i}  \ket{ i } _n \ket{ 1 }.
\end{split}
\end{equation}
After QAE the probability of getting $\ket{ 1 }$ is equal to $\sum_{i=0}^{l_\alpha} \left( \frac{i}{l_\alpha} \right) p_i$
and $\sum_{i=0}^{l_\alpha} p_i = P [x \leq l_\alpha]$. 
Therefore we obtain
\begin{equation}
\frac{l_\alpha}{\mathbb{P}[X \leq l_\alpha]} \sum_{i=0}^{l_\alpha} \left( \frac{i}{l_\alpha} \right) p_i = CVaR_\alpha (X)
\end{equation}
which also achieves quadratic speedup compared to classical Monte Carlo methods.
In our case the random variable $X$ is the delta gross margin $\Delta GM $ defined by \eqref{eq:deltaGM}.

In summary, the three main steps involved in the approach are \textbf{1.} the distribution loading, \textbf{2.} the encoding of the objective function, \textbf{3.} the application of QAE to obtain the relevant statistical quantities. Once we are able to load the data as in Eq.~\eqref{eq:dataloading}, and to encode functions in Eqs.~\eqref{eq:f-VaR} and~\eqref{eq:f-CVaR}, we can efficiently estimate Expectation, VaR and CVaR. The next subsections are devoted to discussing these requirements.

\subsection{Distribution loading }\label{Distribution loading}
One of the main challenges in quantum computation is how to encode the classical data into a quantum computer efficiently. Our goal is to load a random variable $X$  valued in $\lbrace 0, 1, ..., 2^n-1 \rbrace$ with corresponding probability $p_i$ in an $n$-qubit register and form the quantum state
\begin{equation}
\ket{ \psi } _n = \sum_{i=0}^{2^n-1} \sqrt{p_i} \ket{ i } _n.
\end{equation}

Let us collect here some of the most prominent approaches in literature.
The first one is the \textit{qRAM}, namely a quantum version of the random access memory (RAM) \cite{Lloyd}. From a theoretical perspective, it can be described as a quantum operator allowing efficient (i.e., quantum-parallel) access to classically stored information. Unfortunately though such a device was never built at scale.
The second approach is to prepare the desired quantum state directly with \textit{controlled rotation gates} \cite{ Barenco, Kumar, Cortes, Plesch, ventura1999initializing, araujo2021divide, araujo_configurable_2022}. %
The third approach 
resorts to \textit{qGANs} for data loading.
Out of embeddings, qGAN %
encoding stands out due to
its qubit-efficiency and its capacity to be prepared in a polynomial number of gates using qGANs \cite{zoufal2019quantum, herr_anomaly_2020, chakrabarti_quantum_2019, gomez2022towards}. This is a popular choice where parametrised quantum circuit with parameter $\theta$ to generate the quantum state  

\begin{equation}
\ket{ \psi (\theta) } _n = \sum_{i=0}^{2^n-1} \sqrt{p_i (\theta)} \ket{ i } _n.
\end{equation}
In a qGAN the classical generator or discriminator or both is a parametric quantum circuit, thus justifying the term `quantum neural network'.
Despite of low circuit depth and width of the loading circuits, the training of qGANs is critical both in terms of efficiency and accuracy for an arbitrary and large distribution \cite{agliardi_optimal_2022}.

As a consequence, we prefer to ground our experiments on controlled rotations.  %

\subsection{Encoding of the objective function}\label{Amplitude estimation of the payoff function}
Considering the discussion in Subsec.~\ref{subsec: Quantum Amplitude Estimation}, we are interested in piecewise linear functions. Once linear function are available, they can be extended to piecewise linear by resorting to comparators like $F_l$ in Eq.~\eqref{eq:Fl}, that can be easily implemented on quantum computers \cite{draper2006logarithmic, egger_credit_2021}. As for \textit{linear functions}, let $f$ be:
\begin{equation}
f(i) = f_1 i + f_0
\end{equation}
such that 
\begin{equation}
f : \lbrace 0, 1, ..., 2^n-1 \rbrace \rightarrow [0, 1].
\end{equation}
A quantum operator $\mathcal F$ as in Eq.~\eqref{eq:operatorF} corresponding to the linear function $f$ is in general hard to construct, while it is easy to build $\tilde{\mathcal{F}}$ such that
\begin{equation}
\tilde{\mathcal{F}} \ket{ i } _n \ket{ 0 } = \ket{ i } _n \left( \cos [f(i)] \ket{ 0 } + \sin [f(i)] \ket{ 1 } \right).
\end{equation}
Therefore it is now a standard~\cite{stamatopoulos2020option, woerner2019quantum, egger_credit_2021} to Taylor expand $[\sin f(i)]^2$ around a given point, after having introduced a multiplicative constant $C$ in the argument that guarantees the argument is small and therefore the approximation good. Notice that rescaling can be classically accounted as a consequence of the linearity of the involved operations. The approach can also be refined into Fourier expansions of $f$~\cite{herbert_quantum_2022, de_lejarza_quantum_2023}.

\section{Circuit optimization and depth reduction techniques}\label{sec:tricks}
Quantum circuit depth and width minimization as well as gate error mitigation are critical for practical applications of the circuit-based quantum computation. To reduce depth and errors, we follow some hardware optimization techniques, thus getting more accurate result in quantum computers.

\subsection{State-of-the-art techniques}\label{subsec:stateoftheart}
After considering all the different sources of error, we selected couple of error mitigation techniques described in the following. Mapomatic addresses the best qubit mapping for a specific device\cite{mapomatic}; dynamical decopuling takes care of the qubit decoherence; finally we applied mthree to correct measurement errors; and Pauli Twirling to mitigate gate errors. 

\paragraph{Optimal qubit mapping with mapomatic.}
To minimize the gate errors in a specific quantum device for a particular quantum circuit, one of the major concerns is to find the best possible mapping between logical qubits and physical qubits in that quantum device. %
To tackle this issue we use a procedure called \textit{mapomatic} \cite{mapomatic}, which is a post-compilation routine that investigates the best low noise sub-graph corresponding to a quantum circuit in the particular target device. The same quantum circuit is transpiled multiple times, to account for the stochastic behavior of the transpiler optimization. The \textit{mapomatic} ranks the transpiled circuits in terms of the number of the two-qubit gates we target to optimize and finally the best transpiled circuit is picked; for e.g., we choose the circuit with lowest number of swap gates. %

\paragraph{Qubit decoherence suppression with dynamical decoupling (DD).}
The dynamical decoupling 
\cite{dynamicdecoupling} works on physical quantum circuits. %
It performs a circuit-wide scanning for finding the idle periods of time (those containing delay instructions) and inserts a predefined sequence of gates (for example, a pair of $X$ gates) in those spots. The sequence of gates aggregate to the identity, so that it does not alter the logical action of the circuit, but mitigates the decoherence in those idle periods. For the implementation of DD, one needs the duration of each of the instructions natively supported by the backend and has to choose the sequence of the gates, e.g., a pair of $X$ gates. %

\paragraph{Measurement error mitigation with mthree.}
Quantum error mitigation techniques are widely used for reducing (mitigating) the errors that occur during quantum executions. For our project we use \textit{mthree (M3)} \cite{m3errormitigation},
a scalable quantum measurement error mitigation technique that does not need the explicit form of the full assignment matrix ($A$-matrix) or its inverse $A^{-1}$.
Here, the assignment matrix $A$ is defined as the $2^N$ × $2^N$ matrix whose element $A_{i,j}$ is the probability of the bit-string $j$ to be converted into the bit-string $i$ as an effect of the measurement error.
Instead of characterizing the entire $A$, M3 focuses on the subspace defined by the noisy output bit-strings, thus resorting to a number of calibration matrices that scales at most linearly with the number of qubits.

\paragraph{Gate error mitigation with Pauli Twirling (PT).}  In Pauli Twirling (PT), noise channels are approximated as a combination of Pauli gates, so the conjugate of such approximating combination is applied to the circuit to cancel the original noise %
\cite{geller2013efficient, kandala2018extending, cai2019constructing}. PT has become a technique of paramount importance, with implications also in Probabilistic Error Cancellation (PEC)~\cite{PEC2022} and quantum error correction codes~\cite{katabarwa2015logical}. In this work, we follow the results from Ref.~\cite{kim2023scalable}, containing an experimental evidence that twirling the circuits between 1 to 8 times aids in error mitigation.
For our multi-qubit experimental runs performed in Section \ref{Sec: Experimental demonstration}, we opted to twirl one time and used 5000 shots for each circuit run.

\subsection{Approximate Quantum Compiling (AQC)}\label{subsec:aqc}
While computing the relevant statistical quantities using QAE, we encounter quantum circuits with great depth which do not give accurate result on current hardware even after applying the four techniques described above. The general problem of gate compiling is to find an exact representation for a given arbitrary matrix $U\in SU(2^n)$ using the basis set made of CNOT and single-qubit rotations. It is known that the minimum (sufficient) number of CNOTs needed to compile a general gate is $\lceil \tfrac{1}{4}(4^n-3n-1)\rceil$~\cite{shende2004minimal}. 
An extensive research has been conducted in finding precise decomposition which reach this lower bound \cite{khatri2019quantum,cincio2018learning,rakyta2022efficient,rakyta2022approaching,rakyta2022highly}. Such lower bound for exact representation is still computationally expensive in terms of circuit depth due to current chip topology and system coherence times.

Alternative approaches to finding circuit simplifications exist. We are interested in the \emph{approximate} representation, thereby allowing one to move beyond the lower bounds of CNOT cost. Delicate care must be taken not to over-approximate the original circuit. Therefore, we are interested to find an approximate circuit that satisfies a number of hardware constraints (e.g., reduced number of CNOT gates), and is the closest (in some pertinent metric) to the target circuit.  Let $U\in SU(2^n)$ be the unitary matrix induced by the ordered gate sequence of a quantum circuit $\mathcal{S}$ and let $V\in SU(2^n)$ be another unitary matrix, associated to the approximate circuit. Then the pertinent metric is defined as the distance induced by the Frobenius norm, $|| V -U ||_F$. 
Ref.~\cite{AQC} provides a method for finding such approximate circuits with a specified number of CNOT gates, called Approximate Quantum Compiling (AQC). The general AQC Problem is defined as:

\begin{definition}
Given a target special unitary matrix $U\in SU(2^n)$ and a set of constraints for example in terms of qubit topology connectivity, find the closest unitary matrix $V\in \mathcal{V}\subseteq SU(2^n)$, where $\mathcal{V}$ is the set of SU matrices that can be realized by rotations and CNOT gates alone and satisfy those connectivity constraints by solving the following mathematical problem;
\begin{eqnarray}
    \mathbf{AQCP:} \quad \min_{V\in\mathcal{V}\subseteq SU(2^n)} \frac{1}{2} \| V - U \|^2_F\;.
\end{eqnarray}
\end{definition}

\begin{figure*}
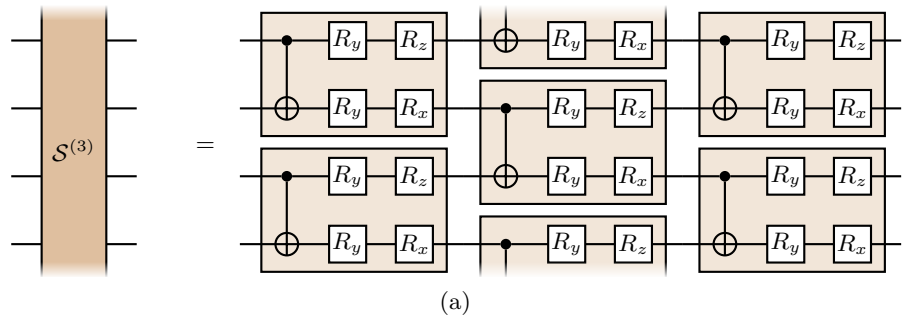
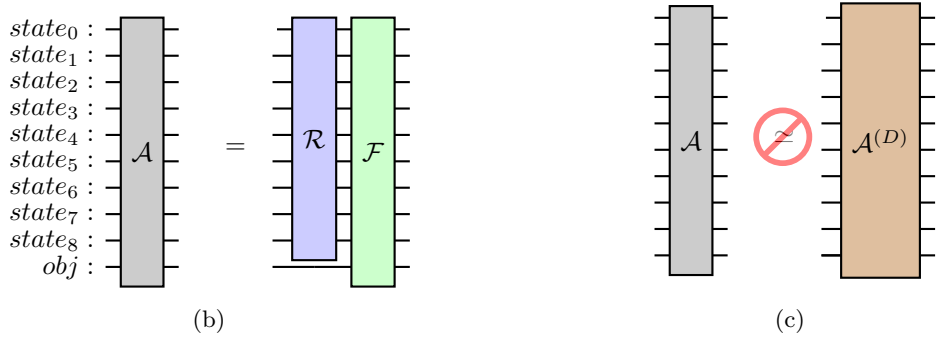
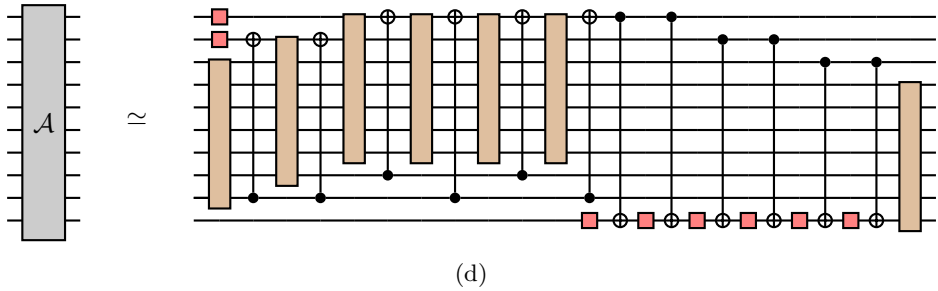
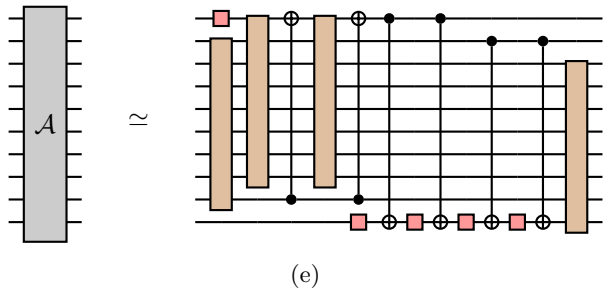

    \centering \small
    \subfigure[]{\input{tex_files/tiles.tex}}\\
    \subfigure[]{\input{tex_files/oracle-0.tex}} \hspace{2.4cm}
    \subfigure[]{\input{tex_files/oracle-1.tex}}\\
    \subfigure[]{\input{tex_files/oracle-2.tex}}\\
    \subfigure[]{\input{tex_files/oracle-3.tex}}
    \caption{
    In the above diagrams a schematic description and workflow of pAQC is presented.  (a) The standard AQC approximates a gate $\mathcal S$ {(the brown block in the LHS)} with an $\mathcal S^{(D)}$ made of $D$ layers ($D=3$ in the picture). Each tile, called a "CNOT-unit", is made of a CNOT gate and four rotations, whose parameters are independent from the other tiles, and determined by the AQC optimization. (b) The operator {$\mathcal{A}$, the grey block} that we want to approximate with pAQC is $\mathcal F \left(\mathcal R \otimes I \right)$, are represented with the green and purple blocks respectively. This is the basic building block for the Grover operator. (c) If the number of qubits is high (10 qubits in the picture), the application of AQC is prohibitive due to the lack of convergence of the optimizer. (d) pAQC is then of help. AQC is applied to each {brown} block, which is identified as to span at most $m=7$ contiguous qubits. The drawing is relative to our specific $\mathcal A$. (e) Same decomposition technique as before, except that the parameter $m$ is now $8$.}
    \label{fig:pAQC_circuits}
\end{figure*}

Given this general problem statement, one can consider so-called "CNOT-units" which are parametrisable 2-qubit circuit blocks consisting of a CNOT gate and pairs of rotations $R_y(\theta_1)$, $R_z(\theta_2)$, $R_y(\theta_3)$, $R_x(\theta_4)$. In this above definition we refer to sets of constraints in terms of qubit topology connectivity. The Qiskit implementation of AQC \cite{qiskit_aqc} has several AQC options including the CNOT-tile network geometry: "sequ", "spin", "cart", "cyclic spin", "cyclic line" and inter-qubit connectivity choices of "full", "line", or "star". Details on such configurations can be found in \cite{AQC} and later in our experimental demonstration we chose the default value of CNOT-tile geometry as "spin" and default qubit topology as "full" (see Figure \ref{fig:pAQC_circuits}(a)).

One can formulate an alternative form of the \textbf{AQCP} problem as finding the closest unitary matrix $V_{\text{ct}}(\pmb{\theta})$ which results from differing products of CNOT-units as depicted in Figure~\ref{fig:pAQC_circuits}(a), where
$\text{ct}$ is a parametrized CNOT connectivity structure and $\pmb{\theta}$ is a vector consisting of all parametrized angles $\theta_i$ from each of the CNOT-units. The problem can then be reduced to finding
\begin{eqnarray}
    \argmax_{\pmb{\theta}} \frac{1}{2^n}|\langle V_{\text{ct}}(\pmb{\theta}), U \rangle|\;.
\end{eqnarray}

The interesting aspect of this problem is that the unitary matrix $V_{\text{ct}}(\pmb{\theta})$ forms a parametric quantum circuit with a \emph{pre-specified} CNOT count. This means that we are able to specify a target CNOT count (alternatively called depth), and find the closest matrix $V_{\text{ct}}(\pmb{\theta})$ to our original target unitary matrix $U\in SU(2^n)$.

This AQC procedure is defined and has been explored for finding approximate quantum circuits close to the target original full quantum circuit. In Section \ref{Sec: Experimental demonstration} we experimentally demonstrate applying AQC to circuits which use up to 8 qubits for computing the Expected Value using the available code as AQC is implemented in Qiskit as a transpiler plugin \cite{qiskit_aqc}. Nonetheless, a natural extension motivated by issues stemming from solving the AQC problem using gradient descent approaches for circuits with a high qubit number, is to segment the target quantum circuit into manageable components and running AQC compilation piecewise before recombining again to obtain the final circuit approximation.

\subsection{Piecewise Approximate Quantum Compiling (pAQC)}\label{subsec:PAQC}

\IncMargin{1em}
\begin{algorithm*}
\SetKwData{Left}{left}
\SetKwData{This}{this}
\SetKwData{Up}{up}
\SetKwFunction{Union}{Union}
\SetKwFunction{FindCompress}{FindCompress}
\SetKwInOut{Input}{Input}\SetKwInOut{Output}{Output}
\SetKwComment{comment}{\#}{}
\Input{Target/original Quantum Circuit $\mathcal{Q}$, maximum subcircuit qubit register span $m$, target individual AQC CNOT depth $D$}
\Output{Approximated Quantum Circuit $\mathcal{Q}_\text{approx}$}
\BlankLine

\textbf{STEP 1:}  Determine subcircuit set $S=\{{\mathcal{Q}_i}\}$ via any block identification technique where each $\mathcal{Q}_i $ act on quantum register sizes at most $m$ qubits \\
\textbf{STEP 2:}  Compute AQC on each subcircuit  \\

    \For{$Q_i$ in \text{S}}{
        calculate the unitary matrix representation $U_i $ of the subcircuit $Q_i$, \\
        compute the AQC approximation $V_i := \text{AQC}(U_i)$ using target depth $D$, \\
        convert the unitary approximation $V_i$ of each subcircuit $Q_i$ back into the AQC CNOT-blocks $\widetilde{Q}_i$
     }
  \textbf{STEP 3:}  Reinsert each newly obtained AQC subcircuit block ${\widetilde{Q}_i}$ into the original quantum circuit (replacing $\mathcal{Q}_i$) to obtain a new complete quantum circuit $\mathcal{Q}_\text{approx}$ which consists of inserted optimized CNOT-blocks via AQC compiling and original circuit components/gates which span qubit sizes on the register greater than $m$ (and hence would affect the convergence of AQC if ran on the full circuit $\mathcal{Q}$ \\  
    \textbf{STEP 4: return $\mathcal{Q}_\text{approx}$ }
 \caption{Piecewise Approximate Quantum Compiling (pAQC) of a quantum circuit}
\end{algorithm*}\label{alg:pAQC}

It is known that AQC compiling fails to converge for high-dimensional $U\in SU(2^n)$ matrices due to Barren Plateau phenomena, and as such, other approaches have been proposed to sidestep the issue \cite{madden2022sketching}. In our case, we propose a technique called Piecewise Approximate Quantum Compiling (pAQC) which applies AQC to appropriately identified subcircuits, here called blocks, see Algorithm~\ref{alg:pAQC}. The central idea of the technique is that instead of approximating the full unitary matrix $U$ representing the quantum algorithm by some `close enough' unitary matrix $V$, we first factor the target matrix into a product of $k$ unitary matrices as $U=U_1U_2\cdots U_k$ such that each unitary matrix $U_i$ is a tensor product defined as $U_i:=W_1\otimes W_2\otimes \dots\otimes W_p$ where each unitary matrix $W_j$ acts on at most $m$ qubits for some suitably chosen size $m$. As we will see, this qubit number $m$ is chosen such that AQC may be applied to each of these subunitary matrices and therefore it cannot be too large e.g. $m\leq 8$. Let $S=\{U_i\}^k_{i=1}$ denote the set of unitary piecewise partitions/factors of the original target unitary. We wish to apply AQC compiling to elements of $S$ (specifically on unitaries $W_j$ which make up each element $U_i\in S$) as $V_i := \text{AQC}(U_i) := \text{AQC}(W_1)\otimes \text{AQC}(W_2) \otimes \dots \otimes \text{AQC}(W_p)$ and then recombine the individual products to create a new approximate unitary representation of the entire target matrix $V' := V_1V_2\dots V_k$. Note that in general, it is not necessary to apply AQC approximation to every element of the tensor product making up each unitary $U_i$ e.g. if $W_1=\text{Id}_2$. 
This procedure is described in Algorithm 1.

Since typical AQC compiling of a target unitary matrix has a target CNOT depth $D$
for a single $\text{SU}(2^n)$ matrix, it is clear that the total circuit depth of a circuit which undergoes $k$-AQC compilations will be greater than or equal to $kD$. Of course, it is also possible to define different strategies where each individual AQC compilation of a circuit slice may have different target depths but for this work we assume they all undergo the same target depth compilation. %

Note that the idea of operating on subcircuits has some commonalities with the `circuit cutting' in literature~\cite{peng2020simulating, tang_cutqc_2021}: in that case, though, subcircuits are chosen to have little interactions and are therefore executed \textit{independently} and then recombined. Compared to such techniques, our objective is simpler and allows for less constrained choices of blocks: we only need to identify subcircuits of $m$ qubits at most, made of adjacent layers in time, on which to apply AQC.
Multiple identification methods are possible, including the cutting techniques in literature, which though would result non-optimal in this context. For simplicity, in our proof of principle, we employ the following rule: we consider the blocks of maximal depth made of at most $m$ \textit{contiguous} qubits. An example for the $9$-qubit quantum circuit for the Expected Value risk measure is depicted in Fig.~\ref{fig:QAE_optimized}, where we compare the pAQC result when choosing circuit cutting span of $m=7$ in Fig.~\ref{fig:QAE_optimized}(d) and $m=8$ qubits in Fig.~\ref{fig:QAE_optimized}(e). 

Our experimental demonstration is shown in Section~\ref{Sec: Experimental demonstration}, for distributions of 9 and 10  qubits, with qubit register size limit $m=7$. %
The number $m=7$ is chosen empirically as the AQC problem is known to have limited convergence for higher qubit counts \cite{madden2022sketching}. Additionally, to keep the total gate depth down to an amount which can justifiably run on IBM Quantum hardware, we chose a target CNOT depth of 10 for each partial AQC compilation.

\section{Dynamic Amplitude Estimation}\label{sec:DAE}

\begin{figure*}[t]
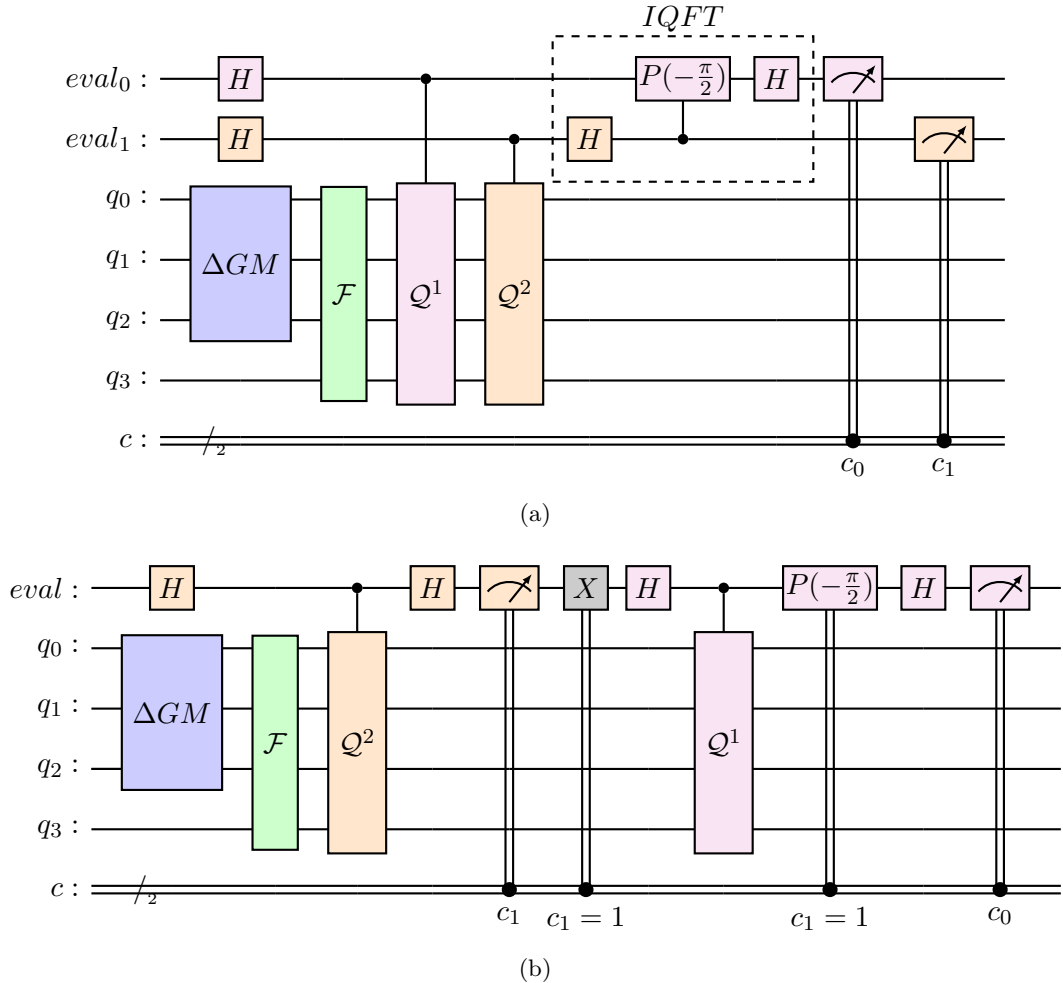

    \centering
    \subfigure[]{\input{tex_files/circuit.tex}}
    \subfigure[]{\input{tex_files/circuit_2.tex}}
    \caption{Quantum circuits for computing $E[\Delta GM]$ with two binary digits precision using  (a) QAE and (b) DAE respectively. Colours help identify the gates relative to the first power (in pink) and the second power (in yellow). In DAE, the controlled $P$ gate is replaced by  a classically conditioned gate through the differed measurement principle. The additional conditioned $X$ gate resets the $eval$ qubit. In QAE two $eval$ qubits used for two binary digit precision (i.e. two iterations), whereas in DAE a single $eval$ qubit is sufficient.}
    \label{fig:DAE_QAE_fig}
\end{figure*}

In this section we describe a modified version of QAE using dynamic circuits, a \textit{Dynamic Amplitude Estimation} (DAE).

In a recent development from IBM Quantum \cite{dynamic_circuits}, a new quantum hardware feature was introduced, namely Dynamic circuits, which incorporate classical processing within the coherence time of the qubits. Dynamic circuits can assist in overcoming some of the limitations of quantum computer. Many critical applications, such as solving linear systems of equations, make use of auxiliary qubits as working space during a computation and features like \textit{mid-circuit measurements}, consisting of the ability to perform quantum measurements while the circuit execution is not yet terminated. The technique was used for instance to validate the state of a quantum computer in the presence of noise, allowing for post-selection of the final measurement outcomes based on the success of one or more sanity checks \cite{mcardle_error-mitigated_2019, shaydulin_error_2021, botelho_error_2022}. Related to mid-circuit measurements, also \textit{mid-resets} were introduced: a qubit, once used, can be returned to the ground state with high-fidelity, so that it is newly available to be used for storage of additional data or for ancillas, thus reducing the total qubit count required in many algorithms \cite{dynamic_circuits}. Physically speaking, a qubit reset is a conditional application of the bit-flip gate $X$: more precisely, a projective measurement of the qubit is performed, and if the result is $1$, the state is flipped from $\ket{1}$ to the $\ket{0}$. If the result is a $0$, nothing is done. Note that a qubit mid-reset implies its mid-measurement. 
The idea of conditioning the application of an $X$ gate to a previous measurement, that is present in the mid-reset, was also extended to any gate or set of gates (sub circuit), giving rise to the so-called \textit{conditional operations}. Many of IBM Quantum’s hardware backends have been upgraded to support dynamic circuits, released in open-qasm3 \cite{10.1145/3505636}. %

\begin{table*}
\centering
\footnotesize
\begin{tabular}{cccccc}
\toprule
Target & Simulator/ & Algorithm & Iterations & Scaled & Unscaled \\ 
statistic & Backend & & & result & result \\ 
\midrule
\multirow{2}*{$E[\Delta GM]$}
& Exact value & Exact & N/A & 8.134 & 10,913,634  \\
& statevector simulator & IAE & 2 & 7.747 & 10,883,563 \\
\midrule
\multirow{2}*{$VaR70[\Delta GM]$}
& Exact value & Exact & N/A & 9.0 & 10,980,925 \\
& statevector simulator & IAE & 2 & 9.0 & 10,980,925 \\
\midrule
\multirow{2}*{$CVaR70[\Delta GM]$}
& Exact value & Exact & N/A & 10.8231 & 11,122,585  \\
& statevector simulator & IAE & 2 & 18.0663 & 11,685,403 \\
\bottomrule
\end{tabular}
\caption{In the above table, the Expectation, $VaR70$, and $CVaR70$ of a 4-qubit $\Delta GM$-distribution are computed exactly and with qasm simulator respectively. By iterations of IAE, we mean the number of quantum-classical interactions. For computing the above statistical quantities with IAE we scale the range of $\Delta GM$-distribution to $[0, 15]$. The corresponding original unscaled values are obtained from the scaled results using Eq. (\ref{scaling_statistics}). 
}
\label{table:4qubitresults}
\end{table*}

The development of DAE involves two main steps. In the first step we replace the usual Quantum Phase Estimation (QPE)~\cite{Kitaev1996QuantumMA} inside QAE with the Iterative Phase Estimation (IPE)~\cite{IPE2007, IAE2022}, thus getting an advantage by reducing the number of qubits, therefore decreasing the costs in terms of noise and hardware requirements. In the second step we further replace the IPE with the Dynamic (Iterative) Phase Estimation (DPE) \cite{dynamic_circuits}, obtaining an advantage over classical quantum feedback/communication time. IPE substitutes the usual Quantum Fourier Transform present in QPE with phase correction terms, which are implemented iteratively and need only a single auxiliary qubit. The accuracy of the algorithm is restricted by the number of iterations rather than the number of counting qubits. IPE builds the phase from the least to the most significant bit. %
In the iteration corresponding to the $k$-th binary digit, the phase corrections are applied according to the outcomes of the previous iterations, that have estimated the digits $k+1,...,m$. This way, the circuit for a given iteration is generated after observing the outcome of the previous iterations.

In DPE, each phase correction term is replaced with a sequence of conditional controlled rotation gates using dynamic circuits. The number of significant binary digits provided by DPE equals the number of iterations.
In Fig.~\ref{fig:DAE_QAE_fig}, we present two schematic diagrams comparing canonical QAE and DAE using dynamic circuits for computing the mean of a 3-qubit gross margin distribution ($E[\Delta GM]$). From the quantum circuits we see that the QPE part in QAE is replaced by DPE in DAE, which reduces the number of evaluation qubits.

Let us emphasize here that, despite QPE being present as a subcircuit in the QAE, the proof of the correctness of QAE does not ground straight-forwardly on that of QPE. In fact, QAE does not produce exactly the hypothesis in which the general theory of QPE holds, and more specifically, the operators $\Delta GM$ and $\mathcal F$ do not load an eigenvector of the associated Grover oracle $\mathcal Q$. On the contrary, they prepare a state which is the weighted sum of two eigenvectors of $\mathcal Q$ corresponding to opposite eigenvalues \cite{rebentrost_quantum_2018}. As a consequence, we cannot directly infer the correctness of DAE from the equivalence between DPE and QPE.

The proof of correctness is anyway very simple, and it grounds on the equivalence between the circuit of QAE and that of DAE. Indeed, referring to Fig.~\ref{fig:DAE_QAE_fig}(a), the controlled powers of $\mathcal Q$ commute, and the blue measurement can be anticipated before the controlled-phase gate transforming into a conditioned gate thanks to the principle of deferred measurement \cite[Ch.~4.4]{Nielsen}, so that it is possible to group all the red gates at the end. The argument easily extends to higher order of powers. From the equivalence of the circuits, it also follows that DAE has the same oracle and query complexity of QAE.

\section{Experimental results}\label{Sec: Experimental demonstration}
In this section, the relevant statistical quantities (mean, VaR, and CVaR) of a $\Delta GM$ distribution are computed, first classically, then implementing QAE, and finally with DAE.

\begin{figure*}
\centering
\input{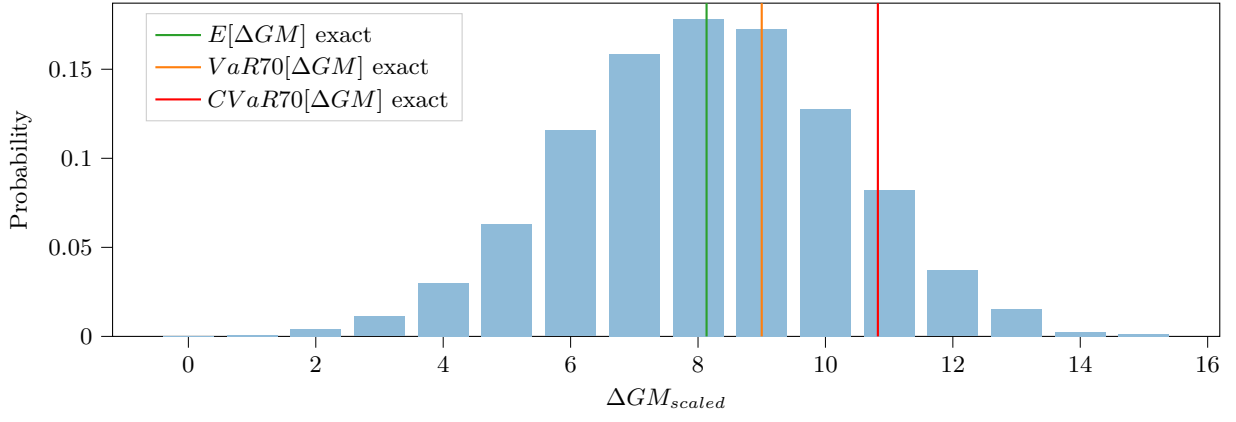}
\caption{The Expectation, $VaR70$, and $CVaR70$ of a 4-qubit $\Delta GM$  distribution are presented. Note that, in the above plot the range of $\Delta GM$ distribution is scaled to $[0, 15]$; therefore the term $\Delta GM_{scaled}$ is used in the X-axis.}
\label{Var_CVar_Classically}
\end{figure*}

\begin{figure*}
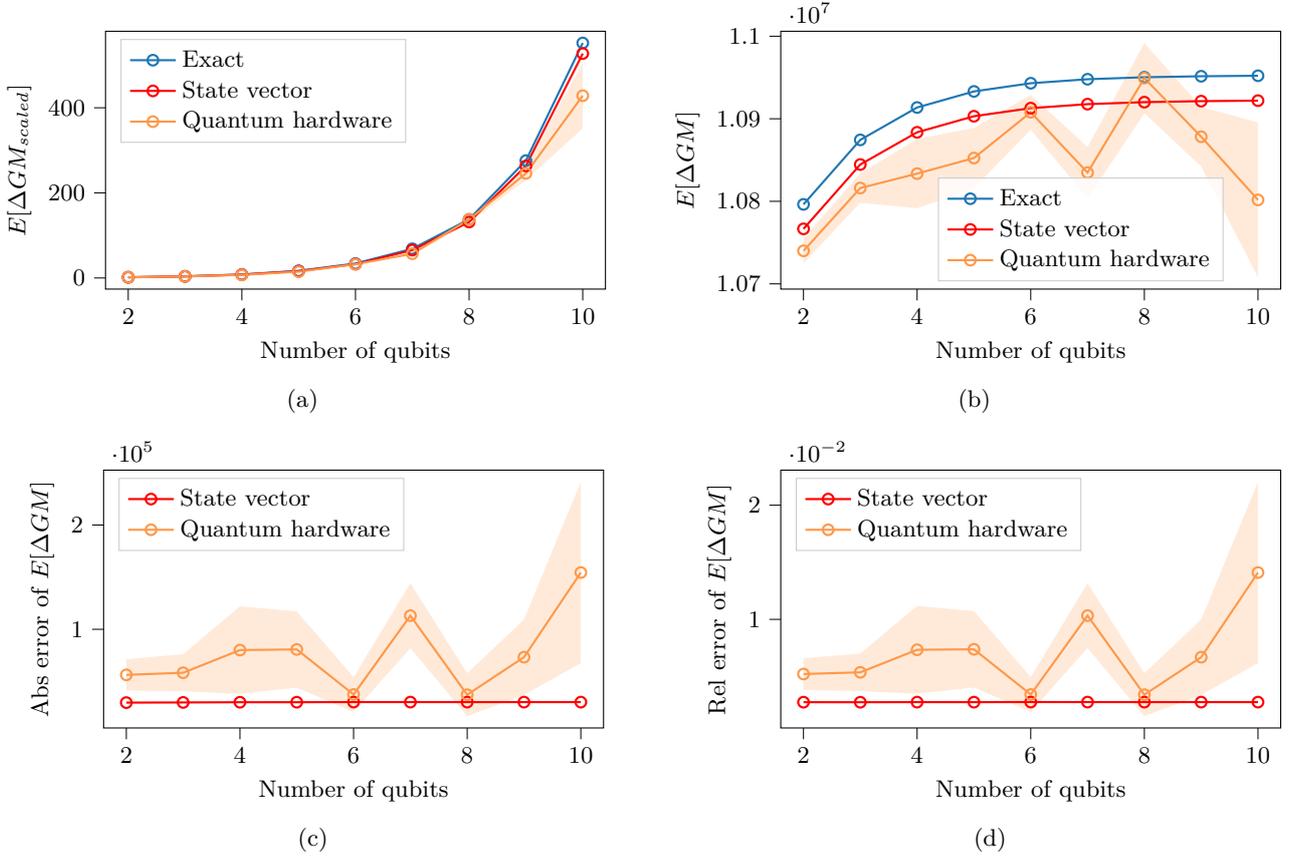

    \centering
    \subfigure[]{\input{tex_files/pic1.tex}}
    \hfill
    \subfigure[]{\input{tex_files/pic2.tex}}
    \\
    \hspace{3mm}\subfigure[]{\input{tex_files/pic3.tex}}
    \hfill
    \subfigure[]{\input{tex_files/pic4.tex}}
    \caption{Experimental results for $E[\Delta GM]$ and the corresponding errors, using IAE and without using error mitigation and circuit optimization, are presented in the above plots. Number of qubits in the X-axes of the above plots are the number of qubits used for the distribution loading, which represents the size of the $\Delta GM$-distribution. In Fig.(a), we plot the Expectations of the $\Delta GM_{scaled}$ against the number of qubits. The term $E[\Delta GM_{scaled}]$ means that the $\Delta GM$ is rescaled to $[0, ..., 2^n-1]$. In Fig.(b), the  $E[\Delta GM]$ values are plotted in its original scale. Fig.(c) represents the absolute errors of the originally-scaled $E[\Delta GM]$, and in Fig.(d) the relative errors of $E[\Delta GM]$ are plotted  as functions of the number of qubits. Relative errors are computed with respect to the exact results. The blue, red and orange curves represent the results obtained from exact computation, state vector simulator and quantum hardware respectively. Each experiment is repeated five times in quantum computer and the shaded regions represent the standard deviations of the quantum computer results around the mean value. %
    }
    \label{fig:QAE_Non_optimized}
\end{figure*}

\begin{figure*}
    \centering
    \subfigure[]{\input{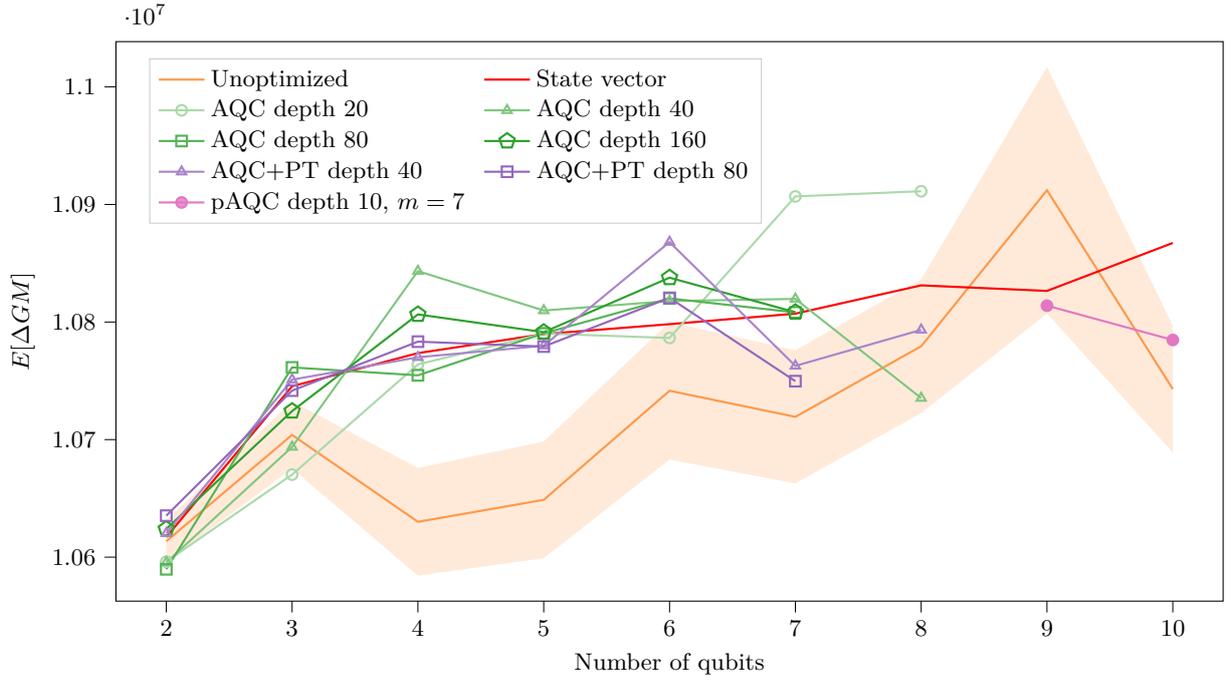}} 
    \subfigure[]{\input{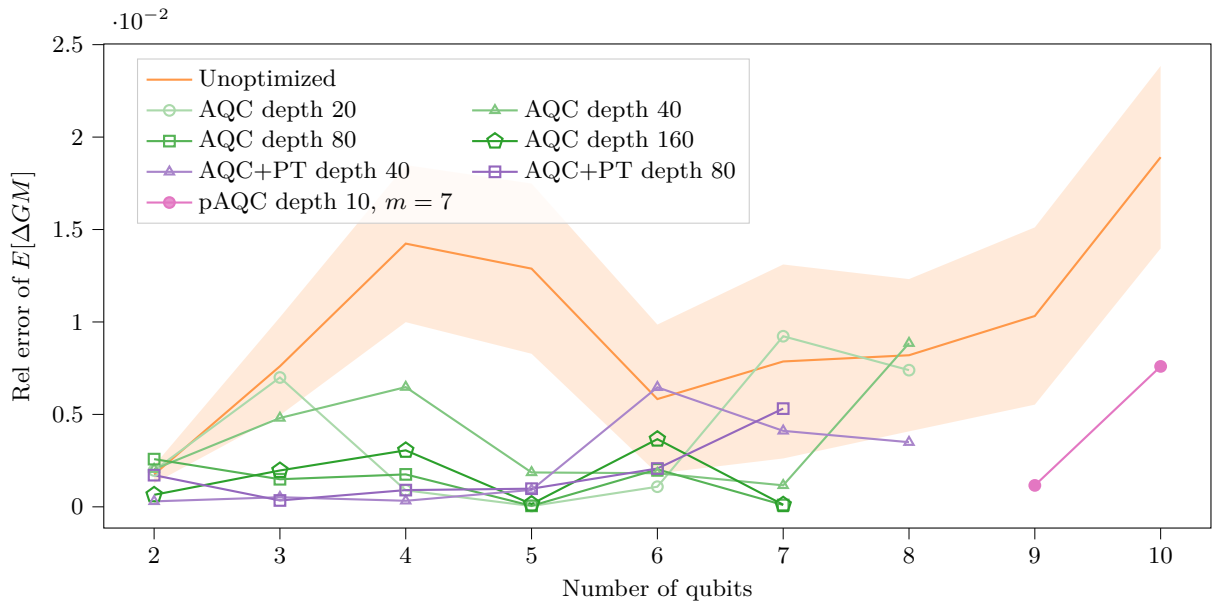}} 
    \caption{Experimental results for $E[\Delta GM]$ and the corresponding percentage errors with IAE using error mitigation and circuit optimization techniques are presented above. In Fig.(a) the $E[\Delta GM]$ values and in Fig.(b) the relative errors of $E[\Delta GM]$ are plotted as a function of the number of qubits. Relative errors are computed with respect to the statevector results. Note that the number of qubits ($n$) in the X-axes of the above plots are the number of qubits used for the $2^n$-dimensional distribution loading; these are not the number of qubits in the quantum circuit after transpilation. The orange and red curves represent the results obtained from the quantum computer and state vector simulator respectively. The other colored lines represent the various types of optimized circuit runs: AQC at target CNOT depth, AQC+PT (Pauli Twiling) represents AQC with additional single PT operation before and after every CNOT gate, and finally, pAQC is used for 9 and 10 qubit data loading where typical AQC-compiling is unstable.
    Each experiment is repeated 5 times in quantum computer. For easier readability, the shaded regions representing the standard deviations of the quantum computer results are shown only around the unoptimized circuit, whereas all other curves and points are the mean values from the 5 experimental runs.
}
    \label{fig:QAE_optimized}
\end{figure*}

\begin{figure*}
    \centering
    \input{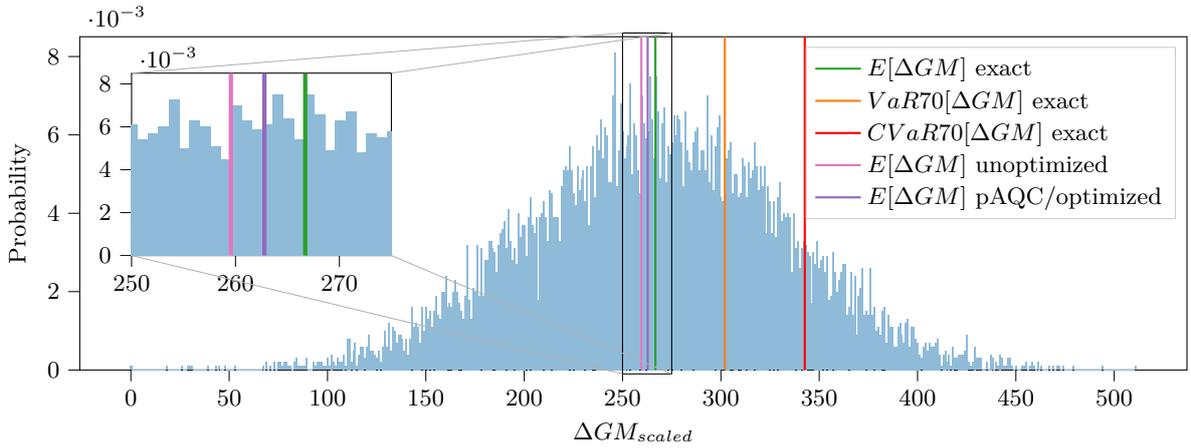}
    \caption{Effect of pAQC/optimized circuit execution on 9-qubit experimental data are shown above. The background distribution represents the complete data set we load into the quantum computer. Coloured vertical lines are the classically computed risk metrics and the calculations of the Expected Value obtained on quantum hardware IBM Mumbai %
    The inset is a zoom of the area around the Expectation, highlighting that the optimization techniques have an improving effect. The detailed hardware results on IBM Mumbai are collected in Figure \ref{fig:QAE_optimized}. For the computational purpose we scale the range of $\Delta GM$ to $[0, ..., 2^9-1]$; therefore, the $E[\Delta GM_{scaled}]$ in the X-axis designates the scaled values of original $\Delta GM$.  %
    }
    \label{fig:pAQC_10qubit_example}
\end{figure*}

\subsection{Input data and preliminary cases}\label{subsec:testing-env}
In this work we consider a simplified weather-related portfolio which depends on gas and temperature. We also consider a one-year time horizon from 1-Jan-2022 to 31-Dec-2022, with daily granularity. The gas prices $[E'_j]$ and the temperatures $[T'_j]$ for $j=0, 1, ..., 364$ are are generated by a two-factor Markov model with $2$ normally distributed random variables, which are assumed to be mutually correlated. We consider $R$ Monte Carlo repetitions to generate random samples of temperatures and gas prices. With the temperature and gas price samples we calculate the change of gross margin ($\Delta GM$), defined in Eq.(\ref{eq:deltaGM}). The generation of $\Delta GM$ is described in our previous work \cite{track_1}.
We prepare 10,000 different values of $\Delta GM $ from a Monte Carlo simulation and a probability distribution with $2^n$ probability values, where $n$ is the number of qubits available for the data encoding. In this first example, we take $n=4$.

The statistical quantities of $\Delta GM$ are calculated on a noiseless simulator as a preliminary test, using IAE.
The 4-qubit probability distribution is loaded in the quantum computer in amplitude encoding, using the procedure described in Ref.~\cite{araujo_configurable_2022}. %
For computational purposes, we need to translate and rescale the range of $\Delta GM$ from $[M_1, M_2]$ to $[0, 2^n-1]$, with $n$ being the number of qubits to encode the distribution. In our case, $M_1= 10,281,599$, $M_2= 11,524,845$ and $n=4$. %
The result is rescaled back with the following formula:
\begin{equation}
\Delta GM = M_1 + \left(\frac{M_2-M_1}{N}\right) \Delta GM_{scaled}.\label{scaling_statistics}
\end{equation}

The results are collected in Table~\ref{table:4qubitresults} together with the exact results of the discretized distribution. %

\subsection{Non-optimized runs on quantum hardware}\label{subsec:test-notricks}
In the previous subsection we have calculated the Expectation, VaR, and CVaR for a random variable with 4-qubit probability distribution using a simulator, with the IAE method. Now we  use quantum computer instead, to compute the Expectation.
We consider the distribution for $\Delta GM$ containing $10,000$ values defined in Subsec.~\ref{subsec:testing-env}, and divide them into $N= 2^n$ bins, where $n= 2, 3, ..., 10$ are the number of desired qubits we want to encode the data into. %

In Fig. \ref{fig:QAE_Non_optimized}, we present the results for $E[\Delta GM]$ using a quantum computer (IBM Kolkata), without applying the optimization methods in Sec.~\ref{sec:tricks}. We used 5000 number of shots for each experiment. For each qubit amount $n= 2, 3, ..., 10$, the experiment is repeated five times, and the mean and standard deviation of their results are visualised. %

\subsection{Effect of circuit optimization and noise mitigation}\label{subsec:test-tricks}

\begin{table*}[htb]
\centering
\footnotesize
\begin{tabular}{cccccc}
\toprule
Target & Simulator/ & Algorithm & Iterations & Scaled & Unscaled \\ 
statistic & Backend & & & result & result \\ 
\midrule
\multirow{8}*{$E[\Delta GM]$}
& Exact value & Exact & N/A & 3.8487 &  10,879,709  \\
& qasm simulator & QAE & 2 & 3.5 & 10,825,519 \\
& qasm simulator & IAE & 2 & 3.6389 & 10,847,105 \\
& qasm simulator & DAE & 2 & 3.5 & 10,825,519 \\
& IBM Mumbai & QAE  & 2 & 3.5 & 10,825,519 \\
& IBM Geneva  & IAE & 2 & 3.5034 & 10,826,048 \\
& IBM Mumbai & DAE & 2 & 3.5 & 10,825,519 \\
\midrule
\multirow{7}*{$VaR70[\Delta GM]$}
& Exact value & Exact & N/A & 4 & 10,903,222 \\
& qasm simulator & QAE & 2 & 3 & 10,747,816 \\
& qasm simulator & IAE & 2 & 4 & 10,903,222 \\
& qasm simulator & DAE & 2 & 3 & 10,747,816 \\
& IBM Geneva & IAE & 2 & 3 & 10,747,816 \\
& IBM Mumbai & IAE & 2 & 7 & 11,369,439 \\
& IBM Mumbai & DAE  & 2 & 3 & 10,747,816 \\
\midrule
\multirow{7}*{$CVaR70[\Delta GM]$}
& Exact value & Exact & N/A & 5.2229 &  11,093,268  \\
& qasm simulator & QAE & 2 & 5.5327 &  11,141,412 \\
& qasm simulator & IAE & 2 & 8.3659 &  11,581,708 \\
& qasm simulator & DAE & 2 & 5.5327 &  11,141,412 \\
& IBM Geneva & QAE & 2 &   9.4449 &   11,749,391 \\
& IBM Mumbai & IAE & 2 & 0  & 10,281,599\\
& IBM Mumbai & DAE & 2 & 9.4449 &  11,749,391 \\ 
\bottomrule
\end{tabular}
\caption{Result summary for computing $E[\Delta GM]$, $VaR70[\Delta GM]$, and $CVaR70[\Delta GM]$ with QAE, IAE and DAE with qasm simulator and quantum computer for $n=3$ qubits. By iterations of IAE, we mean the number of quantum-classical interactions, while for the other methods we mean the number of oracle powers. For the hardware implementation the circuit depth is reduced using AQC. Given the low number of qubits, there is no need for pAQC.  {For computing the above statistic with IAE, QAE, and DAE, we scale the range of $\Delta GM$-distribution to $[0, 7]$. The corresponding unscaled results are obtained using Eq. (\ref{scaling_statistics}).}
}
\label{table:DAE_QAE}
\end{table*}

Using the error mitigation and circuit optimization techniques introduced in Section~\ref{sec:tricks}, we compute the Expected Value risk measure of the $\Delta GM$ distribution for $n=2$ to $10$ qubits. 
Fig.~\ref{fig:QAE_optimized} summarizes the outcomes of such experiments in quantum computer and gives a comparison against the non-optimized case. Each hardware experiment is performed five times and we use 5000 number of shots for each experiment. We implement mapomatic, DD, mthree error mitigation, and PT for 2 to 10 qubit experiments. For circuit-depth reduction we use AQC for 2 to 8 qubit circuits. For 9 and 10 qubit experiments, we use pAQC because from 9 qubit onward the AQC compilation fails due to Barren plateau phenomena. For AQC we experimented with different target depths, for e.g., 20, 40, 80, and 160, and the experimental results are plotted in different colors in symbols in Fig.~\ref{fig:QAE_optimized}. For 40 and 80 AQC depth we compared with and without Pauli Twirling one time, and in general, we see that across the multiple qubit and experimental runs up to 7 qubit data loading size, the Pauli-twirled circuit resulted in more accurate estimates (although within each others variance bounds). For pAQC we use 6 blocks of subcircuits with span $m=7$ qubits and in each block we use AQC with target depth of 10. Since AQC and pAQC approximate the circuit, there is always a \textit{depth-accuracy trade-off}, and the optimal depth can be chosen with extensive empirical trials.  For pAQC also there is a trade off between result-accuracy and span $m$ and number of blocks. In Fig.~\ref{fig:pAQC_10qubit_example}, the relevance of pAQC is explained with a 9-qubit distribution. The  $E[\Delta GM]$ is computed with and without pAQC and we see that applying pAQC we get a more accurate result.

\subsection{Experiments with DAE}\label{subsec:test-dae}

We compare the $E[\Delta GM]$, $VaR70[\Delta GM]$, $CVaR70[\Delta GM]$ computed with QAE, IAE and DAE using Qasm simulator and quantum computer. We use AQC to reduce the circuit depth for real-hardware execution. The result summary is presented in Table \ref{table:DAE_QAE}. Note that, given the low number of qubits, we use directly AQC instead of pAQC to reduce the circuit depth. The result from the quantum computer usually fluctuates from the simulator result due the hardware errors. Therefore, for the consistency purpose we repeat the experiment multiple times in a particular quantum hardware and choose the most probable outcome as the final outcome from the multiple runs. For each experiment we use 5000 shots.

\section{Discussion}\label{Sec:Discussion}

\begin{figure*}
    \centering\footnotesize
    \includegraphics{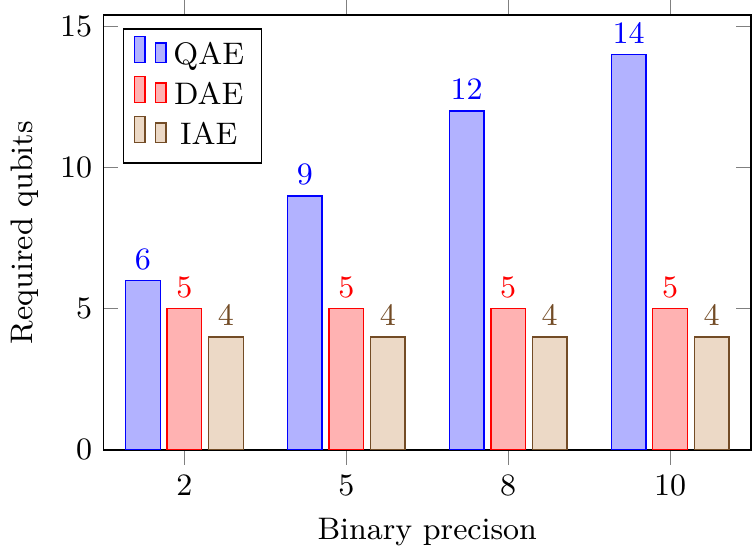}
    \hfill
    \includegraphics{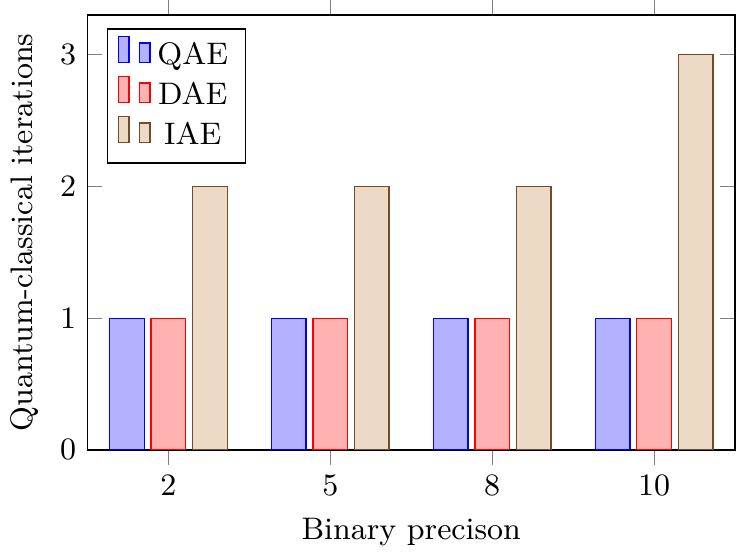}
    \caption{The advantages of DAE over QAE and IAE are represented in the plots above. The binary precision is the number of significant binary digits retrieved for the solution. By increasing the binary precision, on the left we show that DAE uses less qubits than QAE, whereas, on the right the reduction of quantum-classical iterations that DAE brings.}\label{fig:DAE_QAE_and_IAE}
\end{figure*}

\begin{table*}
	\footnotesize\centering
    \begin{tabular}{cccccccc}
    \toprule
    &
        Queries &
        QPE &
        QFT &
        Aux &
        Grover &
        Oracle &
        Generated \\
    &
        &
        &
        &
        qubits &
        oracles &
        powers &
        circuits \\
    \midrule
    QAE \cite{brassard2002quantum} &
        $\mathcal O(\epsilon^{-1})$ &
        Standard &
        Yes &
        $m$ &
        Controlled &
        $2^0,...,2^{m-1}$ &
        Single \\
    MLAE EIS \cite{suzuki_amplitude_2020} &
        Unknown &
        Ad hoc &
        No &
        $1$ &
        Free &
        $2^0,...,2^{m-1}$ &
        MP \\
    QAES \cite{aaronson_quantum_2019} &
        $\mathcal O(\epsilon^{-1})$ &
        Ad hoc &
        No &
        $0$ &
        Free &
        Adaptive &
        MA \\
    IAE \cite{Grinko2021} &
        $\tilde {\mathcal O} (\epsilon^{-1})$ &
        Ad hoc &
        No &
        $0$ &
        Free &
        Adaptive &
        MA \\
    FQAE \cite{nakaji_faster_2020} &
        $\mathcal O(\epsilon^{-1})$ &
        Ad hoc &
        No &
        $2$ &
        Free &
        Adaptive &
        MA \\
    ChebQAE \cite{rall_amplitude_2022} &
        $\mathcal O(\epsilon^{-1})$ &
        Ad hoc &
        No &
        $1$ &
        Free &
        Adaptive &
        MA\\
    \textbf{DAE} (this work) &
        $\mathcal O(\epsilon^{-1})$ &
        DPE &
        No &
        1 &
        Conditioned &
        $2^0,...,2^{m-1}$ &
        Single \\
    \bottomrule
    \end{tabular}
	\caption{A summary of the prior QAE techniques, in comparison with the newly introduced DAE. `Queries' is a theoretically-proven upper bound on the query complexity to achieve a precision $\epsilon$, if available. All techniques are substantially comparable, except for MLAE for which no theoretical bound is known. The notation $\tilde {\mathcal O}$ is used when polylogarithmic factors are discarded. `QPE' is the underlying quantum phase estimation technique: it can be the standard version \cite{Kitaev1996QuantumMA}, the dynamic version (DPE) \cite{IPE2007}, or an ad hoc method. `QFT' indicates whether the Quantum Fourier Transform is applied. `Aux qubits' is the number of qubits required in addition to those needed for the computation of $F$. Here, $m$ is the number of binary digits needed to represent a precision $\epsilon$, namely $2^{-m} = \epsilon$. The original QAE is the only technique whose number of qubits scales with $m$. `Grover oracles' indicates whether the Grover oracles are applied in a quantum-controlled fashion (implying numerous CNOT gates and high associated error), in a classically-conditioned way (implying dynamic circuits and associated circuit delays), or without any control (implying multiple circuits, see last column). `Oracle powers' indicates which oracle powers are evaluated by the algorithm. `Generated circuits' can either be: single (all powers are evaluated through the same circuit), MP (multiple parallelizable, meaning that each oracle power is associated to a different circuit, that can be generated a priori) or MA (multiple adaptive, meaning that each oracle power has an associated circuit, and the power at each iteration is determined after observing the outcome of the previous iteration). In summary, DAE reduces the number of quantum-classical iterations on one side, and of auxiliary qubits on the other.}
	\label{tab:qae}
\end{table*}

In this section we discuss the advantages of our newly developed DAE over usual QAE and IAE along with the error mitigation and circuit optimization techniques we have  implemented.

In line with our objectives, the proposed DAE has the same theoretical time  complexity as the QAE and IAE, providing a quadratic speed up over classical Monte Carlo simulations.
The advantages of DAE {over QAE and IAE} are shown in Fig.~\ref{fig:DAE_QAE_and_IAE}.  {Compared} to QAE, DAE {uses a reduced circuit width:} $\mathcal{O} (n)$ qubits for DAE compare to $\mathcal{O} (n+m)$ for QAE, where $n$ is the number of qubits needed for the objective function and $m$ is the number of significant binary digits required in the {phase} estimation. DAE requires one qubit more than IAE for the evaluation. Since the additional qubit is reset and reused, DAE and IAE have the same 
 {order of} width.
In IAE, {each iteration is} quantum-classical, and each builds a new quantum circuit, based on the outcome of the previous iteration. The number of quantum-classical iterations in IAE is proportional to the precision digits required for the estimation. On the contrary, in DAE all digits are computed in the same circuit.
In Fig.~\ref{fig:DAE_QAE_and_IAE}, we present a graphical comparison of the above advantages of DAE over QAE and IAE. %
Table~\ref{tab:qae} contains a broader comparison of DAE against similar methods.
Empirically, DAE {has proven} to have results in line with QAE on the small problem instances tested in Subsec.~\ref{subsec:test-dae}.

For the circuit optimization and error reduction we used multiple methods. The first technique is the optimal qubit mapping for a specific quantum circuit, which is equivalent to find the best low-noise sub-graph. The second one is the `dynamical decoupling', which inserts identity-summing operations on idle qubits, for mitigating the decoherence.  Next two applications are called the Pauli Twirling and `mthree error mitigation' which is used for the measurement error mitigation. The fifth one is the approximate quantum compiler (AQC). AQC reduces the CNOT depth of the quantum circuit to a specific target CNOT depth by constructing an approximate version of the original circuit. For instance, in our particular three-qubit $\Delta GM$-distribution case the circuit depths for computing $E[\Delta GM]$ with QAE without using AQC is $4130$ (CNOT count: $1464$) whereas with AQC it is $487$ (CNOT count: $154$) respectively. 
Finally, to overcome the limitations of AQC for higher circuit width, we implement Piecewise Approximate Quantum Compiling (pAQC) in Sec. \ref{subsec:PAQC} which applies AQC to appropriate subcircuits. Here we show initial experimental verification that pAQC can be used for 9 and 10 qubit circuits (post transpilation up to 13 physical qubits). However, a deeper experimental analysis concerning the trade-offs between subcircuit indentification rules, individual piecewise AQC depths, and qubit scaling will be part of a follow up work. 
The joint effect of these techniques, discussed in Subsec.~\ref{subsec:test-tricks} and Fig.~\ref{fig:QAE_optimized}, is to roughly halve the percentage of error. Additionally, pAQC can provide results in line with AQC in terms of error reduction, while being applicable to much wider circuits.

\section{Conclusion}\label{Sec: Conclusion}
In this work, a practical implementation of amplitude estimation methods was demonstrated in the field of energy economics. To do so, we first discussed state-of-the-art methods for error mitigation and circuit optimization, designed to reduce the circuit depth and increase the accuracy of the outcomes: specifically, we considered mapomatic for optimal qubit assignment, dynamical decoupling for qubit decoherence suppression, mthree for measurement error mitigation, Pauli Twirling for coherent gate error mitigation, and AQC for depth reduction via circuit approximation. We contributed with pAQC, an improved version of AQC, suited for wide circuits. Thanks to said techniques, we were able to demonstrate an accurate estimation of the Expectation of the delta gross margin of a portfolio, with an input distribution up to 10 qubits. Finally, we developed DAE, a novel amplitude estimation technique, having an lower width than QAE, and reduced quantum-classical communication overhead than iterative methods such as IAE. The new method is also applied to the calculation of $VaR70$, $CVaR70$ of $\Delta GM$. 

We believe this work serves as an important stepping stone in applying quantum risk analysis to real-life data and scenarios on current quantum hardware. The error mitigation techniques used and variants of resource efficient algorithms/compiling demonstrate that by carefully using resources even on today's noisy quantum devices, we are able to run amplitude estimation algorithms with double-digit qubit number on industry relevant use-cases.

\section*{Acknowledgements}
G.A., K.Y., and O.S. acknowledge Travis L. Scholten, Raja Hebbar, and Morgan Delk for helping with the business case analysis; Francois Varchon, Winona Murphy, and Matthew Stypulkoski from the IBM Quantum Support team to help executing the experiments; Jay Gambetta, IBM Quantum for allocating compute time on advanced hardware; Maria Cristina Ferri, Jeannette M. Garcia, Gianmarco Quarti Trevano, Katie Pizzolato, Jae-Eun Park, Heather Higgins, and Saif Rayyan for their support in cross-team collaborations. C.O. and K.G. gratefully acknowledge discussions with Zoltan Zimboras and Peter Rakyta on circuit decompositions; as well as Anton Dekusar and Sergiy Zhuk for useful discussions the Qiskit AQC implementation.

\bibliography{Bibliography.bib} 

\bibliographystyle{quantum}

\onecolumn

\appendix

\section{Hardware specifications of the IBM devices used 
 for our experiments}\label{appendix:device specifications}

In this Appendix we present some of the hardware specifications and calibration details of the IBM devices that were used in this work. We used three quantum devices: IBM Kolkata, IBM Mumbai, IBM Geneva. IBM Geneva has been retired recently, but its specifications can be found in the Appendix~G of Ref.~\cite{eassa2023high}.

\begin{figure}[htb]
\centering
\includegraphics[width=0.9\linewidth]{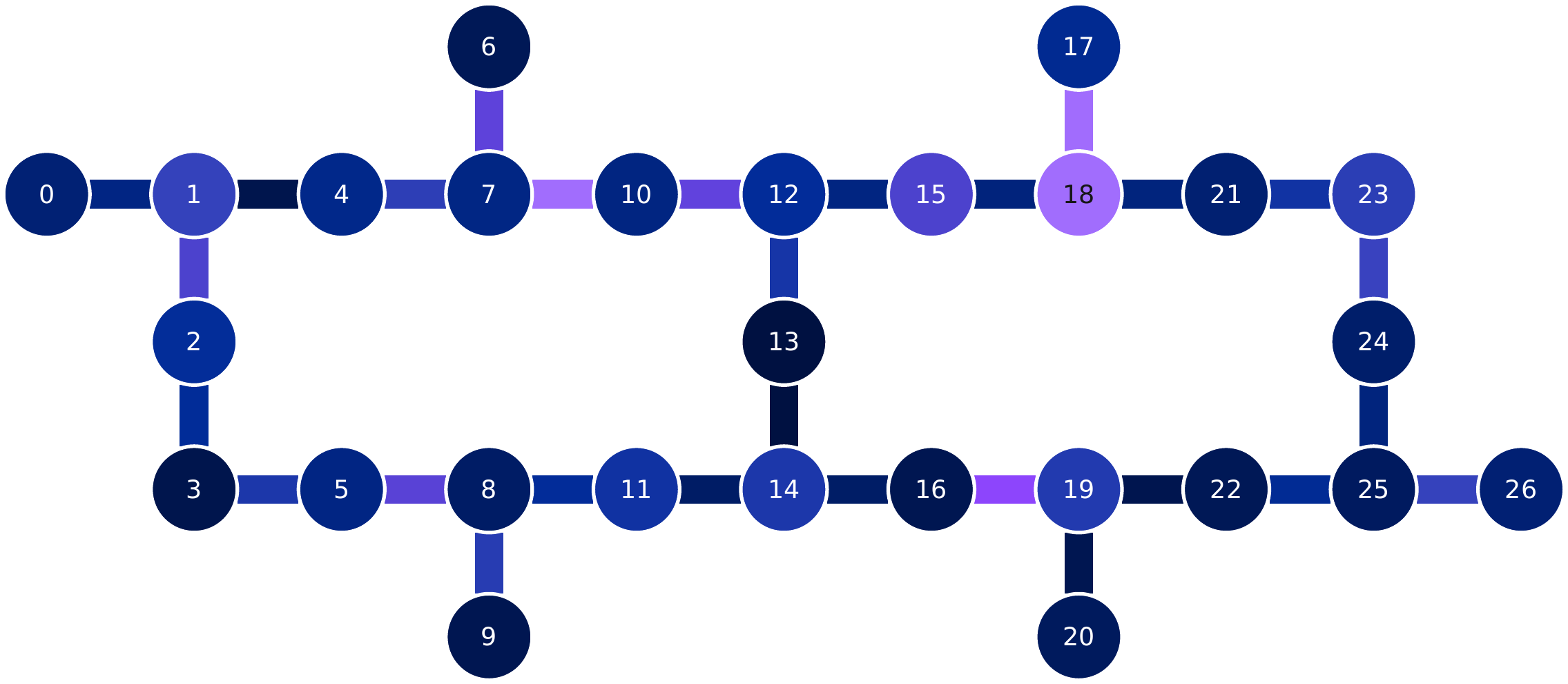}
\caption{The qubit topology graph of IBM Mumbai, which is 27-qubit quantum device.}
\label{fig:mumbai}
\end{figure}

\begin{figure}[htb]
\centering
\includegraphics[width=0.9\linewidth]{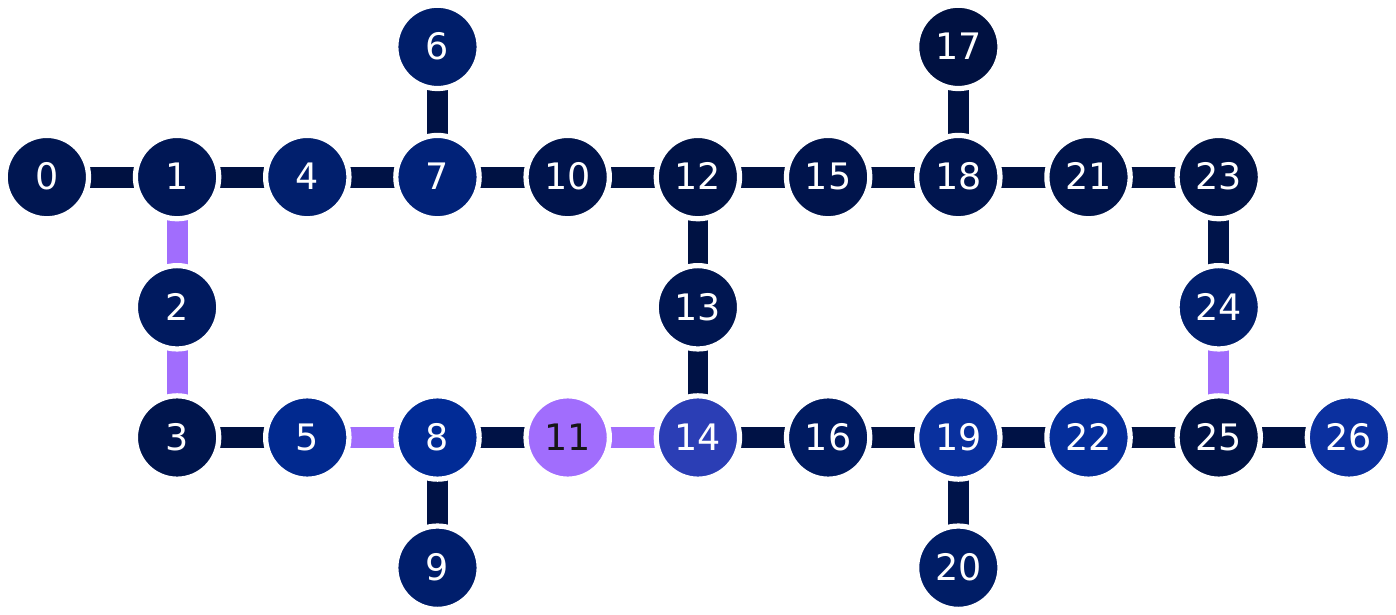}
\caption{The qubit topology graph of IBM Kolkata, which is a 27-qubit quantum device.}
\label{fig:kolkata}
\end{figure}

A topology graph of IBM Mumbai is provided in Fig. \ref{fig:mumbai}, while Table \ref{table:ibm_mumbai} contains some of its specifications. Similarly, Fig. \ref{fig:kolkata} show the topology graph of IBM Kolkata, and Table \ref{table:ibm_kolkata} presents its specifications.

\begin{table}[htb]
\begin{center}
\begin{tabular}{c c c c c c} 
 \toprule
 Qubit &  T1 (us) &  T2 (us) &  Frequency (GHz) &  Anharmonicity (GHz) &  Readout assignment error \\ [0.5ex] 
 \midrule
     0 &  121.893 &   74.716 &            5.204 &               -0.342 &                      0.009 \\
     1 &  184.820 &  193.854 &            4.991 &               -0.345 &                      0.011 \\
     2 &    1.090 &   44.904 &            5.113 &               -0.343 &                      0.015 \\
     3 &  122.597 &   91.154 &            4.866 &               -0.346 &                      0.008 \\
     4 &  137.168 &   97.225 &            5.225 &               -0.341 &                      0.020 \\
     5 &  117.449 &   28.713 &            5.113 &               -0.342 &                      0.033 \\
      6 &  135.955 &   98.217 &            5.203 &               -0.340 &                      0.019 \\
      7 &  117.815 &   46.209 &            5.031 &               -0.346 &                      0.024 \\
     8 &  105.400 &   59.126 &            4.928 &               -0.345 &                      0.036 \\
     9 &  141.072 &  137.121 &            5.054 &               -0.344 &                      0.019 \\
    10 &  101.163 &   41.160 &            5.178 &               -0.342 &                      0.007 \\
    11 &  167.565 &   34.630 &            4.868 &               -0.373 &                      0.137 \\
    12 &   88.162 &  111.563 &            4.961 &               -0.347 &                      0.006 \\
    13 &  141.847 &  289.065 &            5.018 &               -0.346 &                      0.010 \\
    14 &  161.545 &  239.810 &            5.118 &               -0.343 &                      0.064 \\
    15 &  160.563 &  188.404 &            5.041 &               -0.344 &                      0.008 \\
    16 &   72.080 &   67.929 &            5.222 &               -0.340 &                      0.016 \\
    17 &  152.147 &   40.861 &            5.236 &               -0.340 &                      0.004 \\
    18 &  126.501 &   89.403 &            5.097 &               -0.344 &                      0.009 \\
   19 &  116.195 &   28.216 &            5.002 &               -0.345 &                      0.042 \\
   20 &  119.181 &   17.413 &            5.187 &               -0.341 &                      0.017 \\
    21 &  135.261 &   26.209 &            5.274 &               -0.341 &                      0.007 \\
    22 &  136.999 &   40.839 &            5.127 &               -0.343 &                      0.039 \\
    23 &  121.708 &   56.139 &            5.138 &               -0.343 &                      0.005 \\
    24 &  143.578 &   92.588 &            5.005 &               -0.346 &                      0.020 \\
    25 &  231.713 &  119.274 &            4.921 &               -0.347 &                      0.006 \\
   26 &  142.938 &  155.379 &            5.120 &               -0.342 &                      0.043 \\ [1ex] 
 \bottomrule
\end{tabular}
\caption{\label{table:ibm_mumbai} Some of the hardware specifications and calibration details of the {IBM Mumbai} device are collected in the above table. Using this device we produce the experimental results for $E[\Delta GM]$ and the corresponding errors described in Fig. \ref{fig:QAE_Non_optimized}.}
\end{center}
\end{table}

\begin{table}[htb]
\begin{center}
\begin{tabular}{c c c c c c} 
 \toprule
 Qubit &  T1 (us) &  T2 (us) &  Frequency (GHz) &  Anharmonicity (GHz) &  Readout assignment error \\ [0.5ex] 
 \midrule
    0 &   87.465 &  260.251 &            5.071 &               -0.328 &                      0.019 \\
    1 &  154.965 &  253.803 &            4.930 &               -0.331 &                      0.043 \\
     2 &  123.854 &  128.832 &            4.670 &               -0.337 &                      0.027 \\
   3 &  141.994 &  102.277 &            4.889 &               -0.331 &                      0.011 \\
   4 &  138.323 &   59.249 &            5.021 &               -0.330 &                      0.023 \\
   5 &  112.715 &  155.115 &            4.969 &               -0.330 &                      0.022 \\
    6 &  108.421 &   77.457 &            4.966 &               -0.329 &                      0.013 \\
    7 &  230.526 &   45.334 &            4.894 &               -0.331 &                      0.022 \\
    8 &  174.829 &  224.879 &            4.792 &               -0.333 &                      0.016 \\
    9 &   90.581 &  151.520 &            4.955 &               -0.331 &                      0.012 \\
   10 &   83.077 &  116.724 &            4.959 &               -0.331 &                      0.021 \\
   11 &  167.969 &  258.766 &            4.666 &               -0.333 &                      0.031 \\
   12 &  177.792 &  313.115 &            4.743 &               -0.333 &                      0.026 \\
  13 &  171.273 &  255.389 &            4.889 &               -0.328 &                      0.009 \\
  14 &   86.130 &  195.007 &            4.780 &               -0.333 &                      0.035 \\
  15 &  144.229 &   37.756 &            4.858 &               -0.333 &                      0.048 \\
 16 &   88.207 &  181.479 &            4.980 &               -0.330 &                      0.012 \\
  17 &   78.171 &  110.120 &            5.003 &               -0.330 &                      0.024 \\
  18 &   86.443 &  195.045 &            4.781 &               -0.333 &                      0.079 \\
   19 &  249.517 &  298.643 &            4.810 &               -0.332 &                      0.037 \\
   20 &  128.772 &  220.638 &            5.048 &               -0.328 &                      0.014 \\
   21 &  111.219 &  220.139 &            4.943 &               -0.331 &                      0.018 \\
   22 &  163.283 &  224.330 &            4.911 &               -0.332 &                      0.013 \\
   23 &  136.101 &  265.180 &            4.893 &               -0.332 &                      0.040 \\
   24 &  144.812 &   49.376 &            4.671 &               -0.336 &                      0.017 \\
  25 &  190.122 &  144.656 &            4.759 &               -0.334 &                      0.015 \\
  26 &   44.097 &  291.006 &            4.954 &               -0.330 &                      0.019 \\ [1ex] 
 \bottomrule
\end{tabular}
\caption{\label{table:ibm_kolkata} Some of the hardware specifications and calibration details of the {IBM Kolkata} device are collected in the above table. Using this device we produce the experimental results for $E[\Delta GM]$ and the corresponding errors described using the circuit optimization and error mitigation techniques in Fig. \ref{fig:QAE_optimized}. This device is also used to produce the results for $E[\Delta GM]$, $VaR70[\Delta GM]$, and $CVaR70[\Delta GM]$ with QAE, IAE and DAE described in Table \ref{table:DAE_QAE}.
}
\end{center}
\end{table}

\end{document}